\documentclass[12pt,preprint]{aastex}

\newcommand{\et}{et al.}

\newcommand{\ha}{H$\alpha$}
\newcommand{\solar}{\ifmmode_{\sun}\;\else$_{\sun}\;$\fi}

\newcommand{\rha}{$R_{H\alpha}$}
\newcommand{\sigcrit}{$\Sigma_{c}$}
\newcommand{\siggas}{$\Sigma_g$}

\begin{document}

\title{Bright Ultraviolet Regions and Star Formation Characteristics in Nearby Dwarf
Galaxies\footnote{Based on observations made with the NASA {\it Galaxy Evolution Explorer} ({\it GALEX}).
{\it GALEX} is operated for NASA by the California Institute of Technology under
NASA contract NAS5-98034.}}

\author{Nicholas W. Melena\footnote{Current address: The University of Arizona,
1401 E University Boulevard, Tucson, Arizona 85721 USA}}
\affil{Lowell Observatory, 1400 West Mars Hill Road, Flagstaff, Arizona
86001 USA}
\email{nmelena@lowell.edu}

\author{Bruce G. Elmegreen}
\affil{IBM T. J. Watson Research Center, 1101 Kitchawan Road, Yorktown
Heights, New York 10598 USA} \email{bge@us.ibm.com}

\author{Deidre A. Hunter, Lea Zernow\footnote{Current address:
Harvey Mudd College, 301 Platt Boulevard, Claremont, California 91711 USA}}
\affil{Lowell Observatory, 1400 West Mars Hill Road, Flagstaff, Arizona
86001 USA}
\email{dah@lowell.edu, lzernow@hmc.edu}

\begin{abstract}
We compare star formation in the inner and outer disks of 11 dwarf
Irregular galaxies (dIm) within 3.6 Mpc. The regions are identified on
\textit{GALEX} near-UV images, and modeled with UV, optical, and
near-IR colors to determine masses and ages. A few galaxies have made
$10^5-10^6\;M_\odot$ complexes in a starburst phase, while others have
not formed clusters in the last 50 Myrs. 
The maximum region mass correlates with the number of regions as expected from the size-of-sample effect.
We find no radial gradients in
region masses and ages, even beyond the realm of H$\alpha$ emission,
although there is an exponential decrease in the luminosity density and
number density of the regions with radius. H$\alpha$ is apparently
lacking in the outer parts only because nebular emission around massive
stars is too faint to see. The outermost regions for the 5 galaxies
with HI data formed at average gas surface densities of 1.9-5.9
$M_\sun\ $pc$^{-2}$. These densities are at the low end of
commonly-considered thresholds for star formation and imply either that
{\it local} gas densities are higher before star formation begins or
sub-threshold star formation is possible. The first case could be
explained by supernovae triggering and other local processes, while the
second case could be explained by gravitational instabilities with
angular momentum loss in growing condensations. The distribution of
regions on a $\log( {\rm Mass})-\log ( {\rm age})$ plot is examined.
The distribution is usually uniform along $\log ({\rm age})$ for equal
intervals of $\log ({\rm Mass})$ and this implies a region count 
that varies as $1/{\rm age}$. This variation results from either an
individual region mass that varies as $1/{\rm age}$ or a region
disruption probability that varies as $1/{\rm age}$. A correlation
between fading-corrected surface brightness and age suggests the
former. The implied loss of mass is from fading of region envelopes
below the surface brightness limit.
\end{abstract}

\section{Introduction}

Dwarf irregular galaxies are among the most numerous galaxies known and
as such are a fundamental component of the universe. They are the most
common type of galaxy within the Local Group, and are thought to be the
building blocks of larger galaxies. Despite their numbers and proximity
they are still among the least understood of objects. We don't know how
these galaxies form stars or how they have evolved to become the
galaxies we see today. As such, it is essential that we work towards
understanding the forces at work within them, and the processes which
have shaped them.

Understanding the large and small-scale star formation processes in
dwarf galaxies has long posed a problem to the current models. In the
classic Toomre instability (Toomre 1964) there is a specific column
density (\sigcrit ) needed for a ring-like perturbation to grow
exponentially. This model seems to work well in the inner regions of
most spirals (Kennicutt 1989; Martin \& Kennicutt 2001). Dwarf galaxies
are so bereft of gas, however, that in many cases {\sigcrit} is never
reached, even in the centers where the gas column density is highest
(unless the velocity dispersion is very low -- de Blok \& Walter 2006).
Still, we see star formation not only in these centers, but even in the
outer reaches of the disk where the gas density to critical density
ratio (\siggas /\sigcrit $\approx$ 1/Q) is considerably lower than
unity. Similarly, the outer disks of spirals and some entire spiral
disks are subcritical (Martin \& Kennicutt 2001). One possibility is
that localized processes dominate in the absence of large-scale
instabilities (e.g., Elmegreen \& Hunter 2006). Another possibility is
that stars and gas together promote the gravitational instabilities
that lead to star formation (e.g., Yang et al. 2007). A third
possibility is that the gas is unstable even in subthreshold
environments, but the instability growth time is long -- comparable to
the time for angular momentum to leave a growing perturbation through
viscous and magnetic forces (Elmegreen 1987). In all three cases, dwarf
galaxies offer insight into star formation without the influence of
spiral density waves.

In this paper we compare the properties of individual star-forming
regions in outer and inner disks of a sample of dwarf galaxies. The
young regions are identified on {\it{GALEX}} (GALaxy Evolutionary
eXplorer; Martin \et\ 2005) near-UV ($NUV$) images. We use far-UV
($FUV$) and $NUV$, $UBV$, and some $JHK$ colors to model the masses and
ages of the regions. Our investigation includes approximately 2000
separate star-forming regions. With these data we compare mass, age,
and several other key properties of star-forming regions in inner and
outer disks.

Star forming regions in galaxies are also useful for studies of cluster
evolution, particularly regarding fading from stellar evolution and
mass loss, and cluster disruption through cloud collisions and other
processes. An important diagram for these studies is a plot of $\log M$
versus $\log t$ for cluster mass $M$ and age $t$. Such a plot shows a
uniform distribution of points over a range of $\log t$ for each
interval of $\log M$ above the magnitude detection threshold for the
survey (Hunter \et\ 2003; Rafelski \& Zaritsky 2005; Fall, Chandar
\& Whitmore 2005; Bastian et al. 2005a; Whitmore, Chandar, \& Fall
2007). The plot also shows a cluster count decreasing as $1/M$ over a
range of $\log M$ for each interval of $\log t$. The implication of the
former trend is that cluster count $N$ decreases in equal intervals of
age as $dN/dt\propto 1/t$ for all observable masses (Fall et al. 2005).
The implication of the latter is that $N$ decreases in equal intervals
of mass as $dN/dM\propto1/M^2$ for all observable ages (Elmegreen \&
Efremov 1997; Zhang \& Fall 1999; Hunter et al. 2003; de Grijs \&
Anders 2006).  Here we study these diagrams for our sample of galaxies,
finding the same trends, and also study in more detail the distribution
of region surface brightness and radius with age.  

In \S 2 we discuss the images used in our survey. \S 3 describes the
means by which we calculated the photometry for our star-forming
regions, as well as a discussion of our age and mass modeling process.
One of the principle goals of this survey was to examine radial trends
within the dwarf galaxies, and \S 4 discusses these radial trends. 
In \S 5 we present a discussion of some of the major results of our survey:
\S 5.1
considers the surface brightness evolution of regions in four best-case
galaxies, showing a trend with age that is steeper than from stellar
evolution fading alone. This suggests a specific pattern of mass loss
in each region that results from surface brightness limitations. It is
a mass loss that is independent of stellar evolution, evaporation, and
region disruption. Size-of-sample effects regarding surface brightness
are also discussed. 
\S 5.2 reviews the $\log {\rm mass} - \log
{\rm age}$ plots with a comment about the implied time dependence of
region mass, \S 5.3 suggests that outer disk H$\alpha$ is missing
because of faintness, not variations in the initial mass function (see
also Hunter, Elmegreen \& Ludka 2009), \S 5.4 comments on variations in
the star formation rate, \S 5.5 examines outer disk HI column densities
and suggests mechanisms for sub-threshold star formation, and \S 5.6
notes the size-of-sample effects in our observations.  Lastly, \S 6
provides a general overview and summary of what we have learned about
the star formation processes in dwarf galaxies. Also included is an
appendix in which the properties of the individual galaxies within our
survey are discussed.

\section{The Images}

The galaxies in this investigation are a sub-sample of a large
multi-wavelength survey of 136 relatively normal nearby galaxies
without spiral arms (94 dIm, 24 Blue Compact Dwarfs, and 18 Sm
galaxies; Hunter \& Elmegreen 2004, 2006). That survey is
representative of the range in galactic parameters exhibited by dwarfs,
and is aimed at understanding star formation processes in dwarf
galaxies. The $UBVJHK$ and \ha\ data obtained as part of that survey
are described by Hunter and Elmegreen.

To extend this survey into the UV, we obtained images from the {\it
GALEX} archives for 44 of our 136 dwarf galaxies. These data are part
of a study of azimuthally-averaged UV properties in Hunter et al.
(2009). For the present study, we selected 11 dIm
galaxies from this
{\it GALEX} sample, all within 3.6 Mpc of the Sun, to measure the
properties of individual star-forming regions. These data are combined
with comparable measurements on our $UBV$ and \ha\ images and on $JHK$
images of one of the galaxies. The galaxies that are part of the
present study are listed in Table \ref{tab-sample} along with several
basic properties. {\it GALEX} exposure times and tile names are given
in Table \ref{tab-obs}.

{\it GALEX} images were taken simultaneously in two channels: $FUV$,
with a bandpass of 1350--1750 \AA, an effective wavelength of 1516 \AA,
and a resolution of 4.0\arcsec\, and $NUV$ with a bandpass of
1750--2800 \AA, an effective wavelength of 2267 \AA, and a resolution
of 5.6\arcsec. The images were processed through the {\it GALEX}
pipeline, and we retrieved final intensity maps with a 1.5\arcsec\
pixel scale from the archives. The {\it GALEX} field of view is a
circle with 1.2\arcdeg\ diameter; we extracted a portion around our
target galaxies.

We edited foreground stars and background galaxies,
and removed sky from the {\it GALEX} $NUV$ and $FUV$ images.
In some cases the sky was a constant determined from regions around the galaxy,
but in other cases the sky was determined from a low-order,
two-dimensional fit to the surroundings.
We geometrically transformed the UV images to match the orientation and scale of our
$UBVJHK$ images. In that way we could identify regions on the $NUV$ images
and then easily measure fluxes on all of the $UBVJHK$, $FUV$, and
$NUV$ images using the same extraction apertures.

\section{Photometry and Modeling}

We identified individual star-forming regions on the $NUV$ images,
measured their fluxes on all of our broad-band images,
and subtracted the background galaxy light. We
corrected these fluxes to the standard
photometric systems. We then applied Bruzual \& Charlot (2003) stellar
population models to our colors and luminosities to determine the ages
and masses of the star-forming regions. These ages and masses were
studied with respect to galactocentric radius and other variables.

\subsection{Region Selection}

We marked with polygons all of the individual knots of star formation
seen in the $NUV$ images. These regions are defined as spots or
clumps that stand out against the diffuse background disk emission. The
polygons were then applied to all the other image passbands that we
have for each galaxy.

The regions were identified by eye. One source of uncertainty comes
from the occasional difficulty of distinguishing foreground stars from
star-forming regions in the galaxies at the resolution of the $NUV$
images, especially for the more distant galaxies.
Although foreground stars are preferentially red,
we did not wish to bias the regions to only the youngest by selecting only blue objects.
Instead, we limited the sample to galaxies within 3.6 Mpc where the resolution of
the $NUV$ image is about 100 pc.
Thus, most young regions are resolved.

\subsection{Background Subtraction}

The star-forming regions sit on a diffuse disk of galactic UV emission.
We needed to determine the underlying emission for each region in order
to measure the photometric properties of the star-forming sites
themselves. Where we could, we subtracted the mode of an annulus
starting 5 pixels out from the longest radius of the defining polygon
and extending another 5 pixels in radius. 

However, because of the
clustered nature of these sources, this was out of the question for the
majority of them. There were often one or several other regions within
bounds of a reasonable annulus centered on the target region. We
specifically wanted to avoid subtracting one region from another, so in
these cases we found several background spots close to, but separated
from, the star-forming
region and used the median in these regions to characterize the underlying disk.
The separated
background apertures identified on $NUV$ images
were also used as background regions for the other
passbands. The background emission was defined to be the median pixel
value within the background region closest to the star-forming region.
This median pixel value was subtracted from each pixel in the target
star-forming region to get the flux from the star-forming region alone.

To check our determination of the background in the case of the separate
background regions, we determined the sky
using an alternate algorithm (the mode) in one galaxy, DDO 50, using the same background
regions we had used with the original (median) photometry.
The purpose was to ensure that the background value was not biased by the occasional
star that would appear in a background region in the optical that was not there in the UV.
The lack of disagreement or the presence of an
offset between background values determined using the two methods suggest that our background
values in separate regions are robust against this particular problem.

\subsection{Photometry}

After the underlying galactic disk was subtracted, we transformed the
fluxes to the standard {\it GALEX}, $UBV$, and $JHK$\footnote[1]{$JHK$
data were only available for one galaxy in our data set, NGC 2366.
} systems.
The UV data are in AB magnitudes. The $UBV$
photometry is on the Landolt (1992) system, and the $JHK$ fluxes were
calibrated with UKIRT standard stars (see Hawarden \et\ [2001] for a
discussion of near-IR photometric systems). We calculated {M$_{NUV}$},
{M$_V$}, and {M$_J$}\footnotemark[1], as well as the colors $FUV-NUV$,
$B-V$, $U-B$, $J-H$, and $H-K$.
The uncertainties in the magnitudes and colors reflect Poisson statistics in
both the signal and in the sky.

We corrected the photometry for extinction using the same total
reddening E($B-V$)$_t$ as was used in our earlier study of the $UBVJHK$
images (Hunter \& Elmegreen 2006). The total reddening
E($B-V$)$_t$ is foreground E($B-V$)$_f$ from Burstein \& Heiles (1984) given in
Table \ref{tab-sample} plus a constant 0.05 mag for reddening internal
to the dwarf galaxy.
A constant internal reddening of this order for the stars is consistent with measurements
of the Balmer decrement in HII regions in a sample of 39 dIm galaxies (Hunter \& Hoffman 1999).
There the average reddening in HII regions is 0.1, and we have taken half this to represent
the stars outside of HII regions.
We combined the E($B-V$)$_t$ with the reddening
law of Cardelli {\et} (1989) to produce the extinction in each filter.
For the {\it GALEX} filters, $A_{FUV}=8.24{\rm E}(B-V)_t$ and
$A_{NUV}=7.39{\rm E}(B-V)_t$. The photometry for all of the regions are
given in Table \ref{tab-phot}, which is available in its entirety on-line,
and the $JHK$ photometry for the regions in NGC 2366 are given in Table \ref{tab-jhk}.

There are, in total,
1920 regions for all 11
galaxies. There are
$FUV-NUV$ colors for 1913 of them; 1792 of the 1920 have $B-V$ colors
and 43 have $J-H$.
We removed regions from our sample with various problems with the photometry,
including 
photometric errors greater than 1 magnitude, contamination by
foreground stars, unphysical colors, and incomplete photometry.
We were left with 1623 star-forming regions with good photometry.

One reason for the incompleteness of optical and near-IR colors is that
star-forming regions bright in the $NUV$ are not necessarily bright in
the other bands, and can be lost in the galaxy background. Regions
with a few thousand counts in the UV might only have a few hundred
counts in $V$.
In addition, the lower the flux that a region has, the
more susceptible it is to uncertainties in sky subtraction.
For any of our regions
with highly uncertain colors, there was a good probability that our
models would be unable to find a realistic fit. That is why 
we removed all of the star-forming regions
with any photometric errors greater than 1 magnitude.

Another factor working against the smaller regions is that they are
often too faint to be seen at greater distances. While our survey is
complete at a level of $-10$ or $-9$ mag in {M$_V$}, below an {M$_V$}
of $-7$ mag it is incomplete. For the closer galaxies, those less than
1 Mpc away, such as NGC 6822 and IC 1613, this was not a problem, but
at the longer distances of $\ge 3$ Mpc this effect became more
apparent. This problem is illustrated in Figure \ref{fig-disthist},
where the distributions of $M_V$ for all of the star forming regions in
NGC 6822 and DDO 50 are shown. DDO 50 at 3.4 Mpc has a higher
proportion of brighter regions, probably from blending effects, and NGC
6822 at 0.5 Mpc peaks at much fainter levels.

To explore the consequences of distance on our results, we have taken
WLM, which is nominally at a distance of 1 Mpc, and smoothed the images
with a Gaussian kernel to the resolution we would expect at a distance of
3.4 Mpc, the distance of DDO 50. We then analyzed the smoothed images,
including identifying UV-bright regions and background regions,
performing the photometry, and running models. The results are shown
in Figure \ref{fig-wlm} where we compare the region ages and masses
as a function of radius from the original 1 Mpc images and from the pseudo-3.4 Mpc
images.
At 3.4 Mpc, we see fewer regions with more mass and shallower radial gradients, as expected,
but general trends remain. Age-mass ratios, for example, are roughly the same.
From a similar comparison of the other figures that we use in our discussion, we believe that the
results for the more distant galaxies included here, particularly DDO 50 at 3.4 Mpc,
are as valid as those for the closer systems, especially since
our emphasis in this study is on radial trends within galaxies, rather than
differences between galaxies.

\subsection{Mass and Age Modeling}

In order to compare star forming regions, we determine the ages and
masses by finding the best model fit to the colors for each region. The
model colors are from the Bruzual \& Charlot (2003) model stellar
populations with the ``Padova 1994'' stellar evolutionary tracks
(Alongi \et\ 1993, Bressan \et\ 1993, Fagotto \et\ 1994a,b). We use a
Chabrier (2003) stellar initial mass function (IMF). This IMF adopts
the Salpeter (1955) IMF for stellar masses above 1 M\solar. Below 1
M\solar the IMF is shallower than Salpeter's and below 0.3 M\solar, it
flattens out. Mass-to-light ratios are of order 1.4-1.8 times smaller
with a Chabrier IMF compared to a Salpeter IMF. We used the $Z=0.004$
or $Z=0.008$ metallicity models, whichever was closer to the
metallicity expected from nebular oxygen abundances (Hunter \& Hoffman
1999) or, for DDO 210 and LGS3, estimated from the mass-metallicity
relationship. Thus, we used $Z=0.004$ models for all galaxies except
NGC 2366 and NGC 6822. We also use an exponentially decaying star
formation history with a decay time of 1 Gyr for all regions. For the
short ages we obtain here, this means nearly continuous star formation
over the age of the region.

Our modeling program loops over age and finds model colors. For each
model color we subtract the observed color and divide the difference by
the uncertainty in the measured color. We then sum the square of this
difference over the colors. This is a $\chi^2$. Among all possible
solutions, we then average with an exp($-0.5\chi^2$) weighting factor
to obtain a final age and mass. The uncertainties in many of our colors
are sufficiently small compared to the model spectral energy
distribution (SED) fits that the $\chi^2$ values were unreasonably
large (often greater than 40). So, we multiplied all of the color
uncertainties by a constant factor of 5 for all regions in order to get
non-zero Gaussian weights. This is effectively the same as considering
greater than statistical uncertainties, or systematic uncertainties, in
the observations and models. In a small number of cases the $\chi^2$
method does not find a good solution at all. We considered this to be
the case when 5 or fewer of the age models out of the 164 trial ages in
the Bruzual \& Charlot tables with ages larger than 2.3 Myr gave
$\chi^2<40$. We then modified our method to obtain the best possible
SED fit by averaging together the trial ages and masses from only those
age trials that had the lowest rms deviation between the observed and
modeled colors (i.e., not dividing by the measurement errors as in the
$\chi^2$ method). The weights for this average were taken to be
Gaussian functions of this rms deviation, rather than $\chi^2$. The
uncertainties in the results for these cases were taken from the rms
values in the averages.  The final model age, mass, and rms of the fit
are given for each region in Table \ref{tab-phot}.

As a test of the modeling, we tried our model on some well known
clusters: NGC 104 (47 Tuc), a Galactic globular cluster, and NGC 604, a
large young OB association in M33. We obtained $UBV$ colors for NGC
104 from Harris (1996) and measured them for NGC 604  from the Local
Group Survey Data (Massey \et\ 2006).
We measured $FUV-NUV$ for both
clusters from {\it GALEX} archive images. Both sources had their
background light subtracted.  We adopt an E($B-V$) of 0.04
for NGC 104 from Salaris \et\ (2007) and Harris (1996), and use
$Z=0.004$ models, the closest model to the metallicity of the cluster
([Fe/H]$=-0.76$, Harris 1996).
We find an age of $12\pm1$ Gyrs, and a mass
of $(8.9\pm0.9) \times10^5 M_{\sun}$ for NGC 104. Our age
agrees very well with recent estimates: 10--13 Gyrs
(Salaris \et\ 2007), 12.5$\pm$0.5 Gyr (Liu \& Chaboyer 2000),
10.9$\pm$1.4 Gyr (Salaris \et\ 2004), and 11.5$\pm$0.8 Gyr (VandenBerg
2000). Similarly, our mass agrees well with the mass of $7\times10^5
M_{\sun}$ determined by Meylan (1988) from dynamics, surface
photometry, and a two power-law IMF of NGC 104.

For NGC 604, based on the number of observed O stars and a Salpeter
stellar IMF, the mass was previously determined to be $8\times10^4
M_{\sun}$, with an age of 3-5 Myrs (Hunter \et\ 1996). For the models, we use an E($B-V$)
of 0.05, as adopted by Hunter et al., and a Salpeter stellar IMF. The oxygen abundance
of NGC 604 is about half solar (Diaz \et\ 1987, Caffau \et\ 2008), so we use the Bruzual \& Charlot (2003) tables
with 0.4 times solar metallicity.  Then, including the GALEX
$FUV-NUV$ colors, the models give a mass of
$(1.8\pm0.1)\times10^5 M_{\sun}$ and an age of 5 Myrs. These agree with measured
values to within a factor of 2. The model age and mass increase by factors of 2 and 1.3 if
$B-V$ and $U-B$ colors are increased and decreased, respectively, by 0.1 mag to
account for bright emission lines in the $U$ and $V$ bands.

Because our survey was based on $NUV$ images, it was not sensitive to
red, and therefore old, things. It is doubtful that we would see modest
star-forming regions much older than 500 Myrs because of this. There is
also some uncertainty in the models due to the rapidly changing AGB
phases for the age range of 0.3-2 Gyrs (Maraston 2005). While this
problem is more pronounced in the IR than it is in the optical, between
these two problems we feel that our models are reasonably accurate up
to $\sim$500 Myr of age, and this is the age range we emphasize here.

Similarly, we have a lower age limit of 2.3 Myr. This is the point
where the colors begin to change with population age. To start earlier
than this is to say that we can actually see a cluster that is
$10^{5.5}$ yr old, for example, and in fact, such clusters are probably
obscured in the UV. There is no color and stellar evolution before this time
in the models.

One additional issue merits discussion. In modeling star clusters in
the Antennae galaxies, Fall \et\ (2005) found a sharp-edged gap at
ages of 10 Myrs. This is due to the integrated light of massive clusters
at high metallicity being dominated by red supergiant stars.
The cluster colors change so abruptly during this phase (8 - 16 Myrs)
that the model ages become degenerate and avoid ages of 10 Myrs.
This is easily understood in the ``Conti scenario'' of massive
star evolution in which massive stars spend proportionately
less of their He-burning lifetimes as RSGs rather than WRs at higher metallicities
(Massey \& Johnson 1998). On the other hand, we benefit from
the dependence of massive star evolution on metallicity as this effect will be
less pronounced in low metallicity systems
(compare, for example, $Z=0.02$ with $Z=0.004$ in Figures 57 and 59 of Leitherer \et\ 1999).
Furthermore, for Fall \et\ this issue introduced errors of less than 0.4 in $\log t$.
Thus, any gaps we see in the age distributions here are interpreted
as lulls in the cluster formation rate.

\section{Observed Radial Trends}

With this survey we now have masses and ages for 1623 star-forming regions in 11
local group dIms. We also have a rough measure of sizes
from the polygon fits. Over the range of our survey we see ages from 
our lower limit of several
million years to our completeness limit of 500 Myrs,
and masses from millions of solar masses down to just dozens
of solar masses. The individual galaxies are discussed in Appendix \ref{app-indiv}.

We are particularly interested in how different
properties of the star-forming regions vary with radius within each
galaxy, specifically mass, age, size, and number of regions per unit
area. To show a few of these basic properties, for each galaxy we plot
mass against age, mass against galactocentric radius, and age against
radius in Figure \ref{fig-plots}. The figure shows several things: the
relative abundances of star-forming regions at various masses and ages,
the lower mass sensitivity limit for each galaxy as a function of age,
the upper age limit for consistent detections, and the lower age limit
of the models. In the plots of mass or age against radius, it is
possible to see radial trends that may be present.  We also show the
history of star formation in equal intervals of log-age in Figure
\ref{fig-numvsage}. This plots the total number of regions, and the number
of regions more massive than the observational limit at an age of 500
Myr, and the maximum mass of the regions, all as functions
of age.

We are interested in comparing the properties of star-forming regions
in the inner and outer disks. We have made this division in three ways.
First, we cut at the radius \rha\ beyond which \ha\ emission is no
longer detected. In some of our galaxies we find a significant number
of star-forming regions past \rha, and we compare the regions inside
\rha\ with those outside \rha. These are shown as different colors in
Figures \ref{fig-plots} and \ref{fig-numvsage}.
Generally there is little difference in the masses and ages for the regions inside
and outside \rha.

Second, we cut by the number of regions we identified, comparing the
inner 50\% of star forming regions to the outer 50\%. Galaxies with few
UV-selected regions beyond \rha\ are displayed this way in Figure
\ref{fig-plots} as indicated by the panel labels. We show this also for
NGC 6822, DDO 50, and NGC 2366 in Figure \ref{fig-histd50n6822}, where
it is clear for NGC 6822 that there is little difference in the masses
and ages for these two sectors. This is typically the case. DDO 50 and
NGC 2366 differ, however, in that the masses are systematically lower
in the outer regions, as shown in Figure \ref{fig-plots} also.

Finally, we cut at the radius where there is a break in the broad-band
surface brightness profile $R_{Br}$. Three galaxies (DDO 75, DDO 216,
and M81dwA) show double exponential radial profiles: sharp changes in
the rate of decline of the exponential disk seen in broad-band stellar
surface brightness profiles. These breaks are seen in optical and
mid-IR (3.6 and 4.5 $\mu$m) passbands that emphasize all but the
youngest stars and do not represent the ``edge'' of the disk, but
rather a change in the stellar profile. The breaks do not correspond to
a particular surface brightness limit, and their relationship to
identified star-forming regions is inconsistent. The break in DDO 216
occurs well past where we see star-forming regions; in M81dwA there
were only two identified young regions anyway; and in DDO 75 there are
a substantial number of regions past the break. For DDO 75, we compare
the ensemble of regions on either side of the broad-band profile break
in Figure \ref{fig-d75}, showing little difference.

The last two galaxies in Figure
\ref{fig-plots} are plotted without distinction between inner and outer
regions: LGS3 and M81dwA have a few UV regions but no H$\alpha$
regions.

For galaxies with enough star-forming regions, we look at the
luminosity and number of regions per unit area as functions of radius.
These are shown in Figure \ref{fig-lumrad}. The distributions are
similar to each other because the region masses, luminosities, and
sizes do not vary much with radius. Figure \ref{fig-fillingfactor}
shows the summed angular filling factors of the UV star-forming regions
versus radius and Figure \ref{fig-radii} shows the region sizes
(the radius of a circle that has the same area as the polygon that defines the region)
measured on the $NUV$ images. Recall that the sizes are only estimates
from the subjective polygonal outlines of the regions; the filling
factors use these sizes, and are therefore only approximate as well
but for looking for general trends they are adequate.
The figure indicates that filling factors approach $\sim20$\% in the
inner regions. The filling factor profiles are similar to the
luminosity and count density profiles because the sizes are about
constant with radius. Exceptions occur for the few cases where region
size or mass suddenly change at large radii, as for DDO 50, where the
mass decreases and the size increases beyond 4 kpc, as compared to the
rest of the galaxy. In this case the filling factor, which follows the
size and the count density, drops more slowly in the outer part than
the luminosity density, which follows the mass and the count density.
This trend seems to be an exception, however.

The five types of radial profiles---region mass, luminosity density,
count density, filling factor, and size---consistently indicate that the
exponential profile of star formation, and presumably the exponential
disk itself, is the result of an exponentially decreasing count of
star-forming regions, each with about the same mass, luminosity, and
size. The filling factor of star formation decreases too.  The
interpretation is that the basic star-forming unit has constant mass
and luminosity throughout a galaxy disk, even beyond the edge of
visible HII regions, while the star formation rate is proportional to
the number density of these regions, which happens to be proportional
to the local stellar surface density. The reason for the latter
proportionality is unknown, but has the effect of reinforcing the
existing underlying exponential profile with each addition of new
stars.

There is generally a smooth continuation of radial trends beyond
$R_{H\alpha}$, which means that the loss of \ha\ does not
correspond to a change in any physical process related to star
formation. It is more likely an observational artifact, resulting
perhaps from an HII region surface brightness that is below the
detection limit in the outer galaxy. This is to be expected in a low
pressure environment (Elmegreen \& Hunter 2006), because of both the
diffuse nature of ionization bounded HII regions at very low pressure,
and the likely escape of Lyman continuum photons from the star-forming
regions when the surrounding density is low.

\section{Discussion}

\subsection{Surface Brightness Evolution}\label{sb}

Surveys of star-forming regions are limited by surface brightness for
the detection of faint extended objects, rather than by magnitude for
point sources. To understand our detection limits better and to
investigate the evolution of surface brightness, we determined the
average $NUV$ surface brightness of each region, $\mu_{\rm NUV}$, from
the $NUV$ magnitude on the AB scale interpolated to an arcsec
angular size. For resolved regions, this surface brightness is
independent of galactic distance. It has a lower limit that depends on
the exposure time of the image and perhaps the de-projected galactocentric radius
(because of the bias that a selected region must stand out above the
surrounding disk). The lower limit should not depend on age.

Figure \ref{fig-sbvsdist} shows the $NUV$ surface brightness versus
galactocentric distance for four of our galaxies that have a large
sample of regions. The lower limit to surface brightness depends on
exposure time (Table 2): NGC 6822 has over 3.4 times the exposure time
in the $NUV$ than the other galaxies, and a limit that is 
$\sim0.7$ mag arcsec$^{-2}$ fainter. There is no evident
dependence on distance from the galaxy center in 
$\mu_{\rm NUV}$ for NGC 6822, but there is a
small dependence for the other galaxies in the figure. This reflects a
limited bias in the selection of regions for measurement, but should
not be important in studies involving age or mass correlations.

Figure \ref{fig-regsb} shows $\mu_{\rm NUV}$ versus age for the same
four galaxies. Dashed lines indicate the fading trend from the
evolution of a single stellar population model, for which luminosity
decreases with time approximately as $t^{-0.69}$ (\S
\ref{sect:destroy}). Star-forming regions approximately follow the
fading trend down to the surface brightness limit, although the
distribution of points is actually steeper than this for some galaxies.
This steepness suggests either a physical expansion over time for each
region or a loss of mass over time, beyond the loss from stellar
evolution (which is included in the fading trend). The oldest regions
are all above the fading trend because of surface brightness
limitations. That is, many old regions are missing from our survey,
i.e., those below the surface brightness limit on the extrapolated
downward trend in the figure. We have not lost proportionally as many
younger regions by surface brightness fading. The dividing age is
around $10^8$ yrs for this sample. In general, the dividing age between
detection and loss from surface brightness limits depends on the
exposure time and the background noise in the image (in addition to the
local galactic surface brightness, as in Fig. \ref{fig-sbvsdist}).

There is an interesting size-of-sample effect in Figure
\ref{fig-regsb}. The upper limit to the surface brightness decreases
with age much more slowly than the lower limit, indicating that most of
the older regions still observed today were intrinsically brighter than
the younger regions when they formed. Only the faintest observed old
regions, i.e., those at the surface brightness limit, are an
extrapolation of young regions along the fading trend, and then it is
only the brightest young regions that can be extrapolated to fit in
this way. This increase of initial surface brightness with age is a
size-of-sample effect, considering that the sample size of regions in
logarithmic intervals of increasing age is also increasing in
proportion to the age. The older logarithmic bins include more total
regions and therefore a few rare regions of intrinsically large initial
surface brightness. Extrapolated back to their youth, the brightest old
regions today would have had $\mu_{\rm NUV}\sim21$ mag arcsec$^{-2}$ at
an age of $10^7$ yrs. There are no regions in our galaxies as bright as
this today, although the embedded bright source in NGC 2366 would be 
of comparable brightness in the absence of extinction. 

Figure \ref{fig-radiusvsage} shows the region radius versus age for the
same four galaxies as in Figure \ref{fig-regsb}. The radius is about
constant with age, suggesting very little expansion. Considering that
surface brightness is from a combination of mass, luminosity-to-mass
ratio, and radius, this constant radius implies that the steep decrease
in surface brightness with age seen in Figure \ref{fig-regsb} is from a
combination of stellar evolution fading and a loss of stellar mass.
This mass loss must be in addition to supernovae and stellar winds,
i.e., to normal evolutionary effects. The implication is that detected
clusters have lost mass since they formed.

Figure \ref{fig-mnuvvsage} shows the fading-corrected absolute
magnitude of each region in these four galaxies, versus the log of the
age, $t$. Fading correction is done by subtracting $(0.69/0.4)\log t$
from each absolute magnitude. Recall that the luminosity of a region is
proportional to the region mass, $M$, multiplied by the light-to-mass
ratio, and this latter quantity varies with age approximately as ${\rm
t}^{-0.69}$ (\S \ref{sect:destroy}).  The luminosity is also proportional to
$10^{-0.4M_{\rm NUV}}$, so it follows that $-0.4M_{\rm NUV}={\rm
const.}+\log M-0.69\log t$, or
\begin{equation}\left(-1/0.4\right) M+{\rm const.}=
M_{\rm NUV}-\left(0.69/0.4\right)\log t\label{eq:nuv}\end{equation} The
quantity on the right is shown in Figure \ref{fig-mnuvvsage} versus
$\log t$. The dense clusters of points that were sloping downward
steeper than the fading line in Figure \ref{fig-regsb} are now sloping
upward in Figure \ref{fig-mnuvvsage}, which is the direction of fainter
fading-corrected magnitudes over time. A dashed line with a slope of
$+2.5$ guides the eye.  A general slope $\beta$ on this diagram
corresponds to a correlation $\log M={\rm const}-0.4\beta\log t$ from
equation \ref{eq:nuv}. As $\beta\sim2.5$ fits the distribution of
points in the dense portions of Figure \ref{fig-mnuvvsage} for WLM, DDO
50, and perhaps IC 1613, the region masses there seem to be decreasing
as $M\propto t^{-1}$ over and above stellar evolution effects. We do
not see this trend in NGC 6822. This type of mass-age relationship will
also appear in \S \ref{sect:destroy} in reference to the
distribution of points on the $\log M - \log t$ plots in Figure
\ref{fig-plots}, and to the near constant counts of above-threshold
regions in the age histograms of Figure \ref{fig-numvsage}.

Also in Figure \ref{fig-mnuvvsage} there is a long tail of points at
large age with a slight downward slope. These are the overly bright old
regions also discussed for Figure \ref{fig-regsb}. Their distribution
was flat on that diagram because they paralleled the constant surface
brightness limit, and so here they decrease as $(-0.69/0.4)\log t$ by
construction. We are not interested in them for this diagram, but only
in the increasing trend among the denser distribution of points, which
had the puzzling result in Figure \ref{fig-regsb} that it was declining
faster than possible with pure evolutionary fading.

\subsection{Time Dependence of Cluster Mass}
\label{sect:destroy}

Figures \ref{fig-plots} and \ref{fig-numvsage} showed the number of
massive star-forming regions formed as a function of time over the past
500 Myrs. At the upper end of the range of ages in our survey, even the
most massive clusters begin to drop out because of fading. The shift in
minimum detectable mass goes approximately as $M_{fade}=982(t/{\rm
Gyr})^{0.69}\;M_\odot$, so this minimum mass at 500 Myrs is 5 times
larger than at 50 Myrs (Hunter \et\ 2003). Figure \ref{fig-plots} shows
this limit by plotting the minimum mass at 500 Myrs as a horizontal
dashed line, and the slope of the mass-age relationship as a solid
green line. The observed minimum mass for each age is comparable to the
detection limit. The upper end of the green line in Fig.
\ref{fig-plots} is positioned at the intersection of the horizontal
dashed line and the age of 500 Myrs.

Clusters are increasingly lost from our survey over time because of
fading effects as the main sequence turn-off climbs down the mass
function, and because of cluster disruption, stellar evaporation from
the cluster, and surface brightness limits. Piskunov \et\ (2006) find
the average open cluster lifetime in the Milky Way to be $322 \pm 31$
Myrs. Boutloukos \& Lamers (2003) and Gieles et al. (2006) suggest that
the disruption time for a $10^4$ M\solar cluster depends on the
galactic environment, including the tidal field, molecular cloud
collisions, and shear rate. Boutloukos and Lamers find that the
disruption time in the SMC is very long---8 Gyr, much longer than the
time scale they derive for spiral galaxies, and longer than the time
derived by Piskunov et al. for the Milky Way. Hunter \et\ (2003) also
suggest that clusters in the LMC and SMC have long disruption times.

The number of clusters per unit age in each mass interval decreases
like $\sim1/{\rm age}$, which in a mass-age diagram like Figure
\ref{fig-plots}, means that the density of points is approximately
constant along horizontal strips (Fall, Chandar, \& Whitmore 2005).
This constancy is also evident in Figure \ref{fig-numvsage} as a nearly 
flat histogram of cluster counts above the threshold mass limit.
Such a time depletion implies that clusters are losing mass or getting
disrupted constantly. In addition, the mass function of clusters in the
LMC remains a power law for all observed times (Elmegreen \& Efremov
1997; Hunter et al. 2003), with no preferred loss of low mass clusters
in older age groups. In the case where each cluster loses mass like
$M\propto 1/t$, which is one way to give the observed distribution on
the mass-age plane, the lack of a low mass turnover in the cluster mass
function implies that the lowest mass clusters are lost from the survey
because of fading limitations, and not disruption. Because the galaxies
in the present survey are dwarf irregulars with internal tidal forces
like those in the SMC and LMC, and because brightness limits are higher
for the present survey than for the LMC and SMC, we expect that fading
is the dominant cause for the loss of low mass clusters in the present
survey too. If physical disruption were the cause instead, then there
would be a gap between the lowest mass cluster for each age on the
mass-age plane, and the fading limit. There is no such gap in Figure
\ref{fig-plots}.

In another interpretation of cluster evolution, clusters keep their
mass for a time $t$ and then get disrupted suddenly, perhaps in a GMC
collision. In this case, the disruption probability per unit time
should decrease with time as $1/t$ to get a constant number per unit
$\log t$ in fixed intervals of mass, as suggested by the distribution
of points on the mass$-$age plane.

Our observations cannot distinguish between these two mechanisms of
physical cluster loss, i.e., between mass loss from each cluster with
cluster mass $\propto1/t$, or sudden cluster loss with a disruption
rate $\propto1/t$. In both cases, the loss mechanisms have to be
independent of cluster mass if the mass function has the same power law
slope for all ages. The first mechanism cannot be simple evaporation
because evaporation gives a constant cluster loss rate,
$dM/dt=$constant for a constant cluster density (as determined by tidal
effects, for example; Baumgardt \& Makino 2003). To have $M\propto1/t$
for each cluster, we need an evaporation rate $dM/dt\propto1/t^2$ or
$dM/dt\propto M^2$, which is not the way evaporation works.  The second
mechanism cannot be simply cloud-cluster collisions in the usual model
because the survival probability in a random ISM scales with time as
$\exp(-t)$, from Poisson statistics.

The discussion in \S \ref{sb} suggests that some individual region
masses decrease with time as $1/t$. There may be some sudden disruption
of regions too, with probability $\propto 1/t$, but we cannot see that
in our data (we know nothing about when the regions that have been
disrupted were disrupted).

A schematic model for mass loss by fading below the surface brightness
limit of a survey is shown in Figure \ref{fig-massloss_sblimit}. The
two Gaussian curves represent projected profiles $I$ through two
clusters having the same total mass. One cluster is bright and the
other is dim and slightly larger in radius. The horizontal red line is
the surface brightness limit, which is above the outer envelopes of the
clusters. The green curves are the running masses measured only above
the surface brightness limit. They increase from left to right as the
cumulative cluster masses increase from left to right in the projected
images. These running masses are given by $\int_{x_0}^x \pi x
I^\prime(x) dx$ for position $x$ and starting position $x_0$ on the
left in the diagram. The intensity in the integral, $I^\prime$ is taken
to be zero for the parts of the cluster below the surface brightness
limit, and the actual cluster intensity $I$ for parts of the cluster
above the surface brightness limit. The dispersion of the bright
cluster is 1 on this scale, and the dispersion of the dim cluster is
$1.682$, designed to make the two clusters have the same size out to
the surface brightness limit. This satisfies the constraint suggested
by Figure \ref{fig-radiusvsage} that clusters do not appear to expand
much with age. In fact, if we identify the faint cluster in Figure
\ref{fig-massloss_sblimit} as an old cluster, then it has expanded with
age, but at the same time as it has faded, leaving the apparent size at
the surface brightness limit about constant.  The point of Figure
\ref{fig-massloss_sblimit} is that two clusters with the same total
mass can have two different measured masses if only those parts of the
cluster that are above the surface brightness limit are measured.
Cluster fading then automatically causes a measured mass loss over and
above stellar evolution effects. This mass loss is not the same as
stellar evaporation because the missing stars are still near the
cluster and maybe even bound to the cluster. They are just in the
faint, unobserved, envelope.

Several of our galaxies have peaks in the number of star-forming
regions as a function of time, as well as peaks in the number of
massive regions and in the maximum mass of clusters. From a study of
integrated \ha\ equivalent widths of a large sample of dwarfs, Lee \et\
(2007) conclude that dwarf galaxies should experience 1-3 starbursts
per Gyr. These bursts should have a duration of 50-100 Myrs and an
amplitude of 6-10 times the star formation rate of the quiescent
period.
McQuinn \et\ (2009) have found large-scale bursts
of duration 200--400 Myrs in several nearby dwarfs, with smaller
scale ``flickering'' on timescales of 3--10 Myrs.
In our sample we see two gaps of duration 60 Myrs and 1300 Myrs over the
past 1 Gyr in NGC 6822. The enhanced cluster formation episode centered on
an age of 1 Gyr, for example, lasted about 470 Myrs.
Thus, the history of cluster formation in NGC 6822 has been roughly
consistent with the findings of these studies.
However, in 6 other systems in our sample, cluster formation appears
to have been roughly constant until 9--50 Myrs ago, with few or none
formed since then. Three galaxies show no obvious gaps in
cluster formation over the past 500 Myr.

\subsection{Comparing Inner and Outer Galactic Disks: The Lack of H$\alpha$}

We see no consistent change with radius in the average mass or size of
star-forming regions for each galaxy. Regions beyond $R_{H\alpha}$ have
masses as high as (1.8 $\pm 0.9)\times10^5 M_{\sun}$ for the globular
cluster in WLM, and (3$\pm 1)\times 10^5 M_{\sun}$ beyond $R_{H\alpha}$
in NGC 6822. This lack of a mass gradient is consistent with the lack
of any changes in the UV radial surface brightness profiles at the
point where \ha\ ends (Boissier \et\ 2007). Exceptions are DDO 50 and
NGC 2366, which show the clearest gradients in region mass. DDO 50 and
NGC 2366 are compared to a more standard case, NGC 6822, in Figure
\ref{fig-histd50n6822}. Mass and age histograms are plotted for each
galaxy, with the inner and outer half of the regions separated for
comparison.  The median log (mass) for NGC 6822 is only 0.12 larger in
the inner half than the outer half, while the median log mass is 0.25
larger in the inner parts of DDO 50 and NGC 2366 than the outer parts.
The median ages are slightly younger in NGC 6822 and NGC 2366 and
slightly older in DDO 50 in the outer halves.

H$\alpha$ becomes invisible in the outer parts of galaxy disks even
though we find normal star formation there. For DDO 210, there is no
H$\alpha$ emission even though there is one region younger than 10 Myr.
There are many reasons why H$\alpha$ should disappear when star
formation continues, including a lack of massive stars due to old
cluster age or a steep IMF (see, for example, Pflamm-Altenburg
\& Kroupa 2009), an escape of Lyman continuum photons from
the vicinities of the O-type stars, and a undetectable surface
brightness for the ionization-bounded HII region. The age gradient
explanation is difficult to reconcile with our observation of a nearly
constant age with radius, and that, along with possible IMF variations,
are also in conflict with the observation of a normal exponential UV
profile. The UV exponential implies that the onset of star formation in
the outer disk is in the same proportion to the surrounding stars as in
the inner disk, and in a steady state, this implies that the evolution
of O-stars off the main sequence is in the same proportion too. Thus
the fraction of star-forming regions in each stage of evolution should
be independent of radius, as should the IMF, and this means that the
fraction that produce H$\alpha$ is constant too. An exception might
occur if the outer regions had a burst of star formation at just the
right time in the past to have all of the associated O-type stars now
evolved off the main sequence. This is unlikely for all of the galaxies
here.

Escape of Lyman continuum radiation from the vicinity of the O-star can
explain the lack of H$\alpha$ whether or not those photons escape the
galaxy. H$\alpha$ emission disappears when the ionizing radiation from
a massive star spreads out over a large volume and the emission measure
becomes undetectable. Considering that the quantity
$n_e^{2/3}R_e\sim50$ cm$^{-2}$ pc for an O-type star (Vacca et al.
1996) with HII region electron density $n_e$ and radius $R_e$, and that
the density in the ambient interstellar medium should be $0.1$
cm$^{-3}$ or lower in the outer regions of most galaxies, a typical HII
region radius would be $R_e>180$ pc and the corresponding emission
measure would be $n_e^2R_e<1.8$ cm$^{-6}$ pc, which is undetectable in
our survey by a factor of 3--40. The limit of our detection for emission
measure is 5.3--74 pc cm$^{-6}$ (Youngblood \& Hunter 1999),
which corresponds to a minimum density of 0.4--1.4 cm$^{-3}$ at this
ionization rate. Lower densities, particularly off the midplane, would
make $R_e$ even larger than several hundred pc, and then the Lyman
continuum photons could escape the galaxy completely.

\subsection{Trends in Recent Cluster Formation Histories}

Our dIm galaxies fall into one of three categories describing the
overall recent star formation histories of the past 500 Myrs:

{\it Star formation with gaps}.
These galaxies take two forms.
First, NGC 6822 is an example of periodic
spikes in star formation over a finite duration, and it has made
regions as massive as $\sim3\times10^5\;M_\odot$.
The most recent gap in cluster formation occurred around 100 Myrs ago
and lasted about 60 Myrs.
The second form are those galaxies with an apparent cessation in
cluster formation in recent times. These galaxies include
DDO 50, DDO 70, DDO 216, and WLM. The
recent gap began 9 (DDO 60 and DDO 75) to 25 (WLM) Myrs ago.
DDO 70 and DDO 75 sit in between these two types:
They appear to have had a fairly constant level of star formation
for the last several hundred million years until about
10 Myrs ago. In the past 10 Myrs only a few clusters have formed.
In addition, clusters older than about 200 Myrs are missing
except for a few especially massive ones: clusters up to $4\times10^4$ M\solar\
in mass.

{\it Constant star formation}. These galaxies appear to have had a
reasonably steady, nonzero rate of star formation over the last 500
Myrs. They make a $5 \times 10^3 M_{\sun}$ region on the order of once
every 10 Myrs. They make a $5 \times 10^4 M_{\sun}$ mass cluster about
once every 100 Myrs, and some have managed a $10^5 M_{\sun}$ region at
some time in their recent past.
Galaxies in this category include IC 1613 and NGC 2366.

{\it Nearly no star formation}.
These galaxies
show up as little smudges barely above the background sky on our
\textit{GALEX} images. LGS3 is our archetype for this category, with
only 6 useable regions in the last Gyr, all only a few 100 M\solar.
We also include DDO 210,
as we found only 9 useable regions, none of which were more massive
than $\sim10^4 M_{\sun}$; none of the 5 from the last 500 Myrs was
more massive than $10^3$ $M_{\sun}$. We also include M81dwA in this
category; this galaxy has only two measured regions.
These three galaxies are also the only galaxies in our survey that
show no measurable HII regions.

\subsection{Gas Densities in the Outer Disks: Sub-threshold Star Formation}

We examined the HI gas density at the star forming region that is
furthest from the center for each of the five galaxies with HI maps.
Using gas surface density radial profiles, we calculated the azimuthally-averaged HI density in
$M_\odot\ {\rm pc}^{-2}$. These values are presented in Table
\ref{tab-param}. There is a range in gas surface density at the radius
of the farthest star-forming region, $\Sigma_{gas}(R_{UV})$, from
1.4--4.4 (1.9--5.9, corrected for He) $M_\odot\ {\rm pc}^{-2}$. Schaye
(2004) suggests that the threshold for star formation is a surface
density of 3--10$\times 10^{20}$ HI atoms cm$^{-2}$, or roughly 2.4--8
$M_\odot\ {\rm pc}^{-2}$. Hunter \et\ (2001) find that 2--8 $M_\odot\
{\rm pc}^{-2}$ is required for star formation as defined by HII regions
in NGC 2366. The values of $\Sigma_{gas}(R_{UV})$ found here are at
best at the low end of these previous limits in three of the galaxies.
The furthest star-forming regions are not incredibly massive, but high
masses are not expected for single-member populations (i.e., furthest
regions), according to the size of sample effect (\S
\ref{sect:sample}). Still, there is star formation at low average gas
column densities. One possibility is that there is a large amount of
cold molecular material in outer dwarf disks. Galliano \et (2003) find
that in NGC 1569 there is a significant millimeter excess, perhaps
indicating extensive cold (5-7 K) dust that may make up 40-70\% of the
total dust mass of the galaxy. If this dust is real, then this is
perhaps evidence of pervasive cold molecular material. Alternatively
and more likely, the star-forming regions may be at places where the
{\it local} gas density is significantly higher than the average
surroundings (Schaye 2004; Elmegreen \& Hunter 2006). The problem then
is to determine the mechanism of cloud formation in sub-threshold gas.

Most of the far outer star-forming regions in spiral galaxies shown by
Thilker et al. (2007) are in spiral arms. Gaseous spirals can propagate
much further out in a disk than stellar spirals because the gas spirals
do not get absorbed at the outer Lindblad resonance like the stellar
spirals do. Thus there is a natural explanation for cloud formation in
outer spiral disks that is connected with compression from gaseous
density waves.  The far-outer regions in dwarf Irregular galaxies are
not aligned in spiral arms, however, so this explanation does not apply
here. There has to be another source of compression for Irregulars that
is consistent with the continuation of the exponential disk. This
exponential constraint would seem to rule out random extragalactic
cloud impacts or minor mergers, which would trigger star formation in a
more irregular way.  There is also little evidence for supersonic
turbulence in far-outer disks, so the compression from that would seem
to be small. For example, the HI line widths in outer disks could be
largely thermal at several thousand degrees Kelvin, i.e., from a nearly
pure warm phase (Young \& Lo 1996, Young \et\ 2003, de Blok \& Walter 2006).

An interesting possibility is that star formation is triggered by
gravitational instabilities in the outer disk, just like it is thought
to be triggered in the inner disk.  This can occur in subthreshold gas
if the angular momentum from a growing perturbation is removed. The
Toomre Q threshold applies only when perturbations conserve angular
momentum (Elmegreen 1987). With magnetic tension or viscosity that can
remove angular momentum from a growing perturbation as it tries to spin
up, the disk is always unstable even at large $Q$. What changes is the
growth time, which becomes long at large $Q$, in rough proportion to
the viscous time or magnetic damping time (Elmegreen 1991; Gammie
1996). If the viscous and damping times are approximately equal to the
crossing time through the thickness of the disk, which is true for
equipartition magnetic fields and turbulent damping, then there should
be no sharp transition at the $Q=1$ radius. The instability growth time
would increase steadily with radius in proportion to the disk thickness
crossing time, and this would always give some level of instability for
cloud formation.  Previous observations of a threshold, based on a drop
in H$\alpha$ emission, would then be an artifact of low emission
measure and Lyman continuum photon escape, as discussed above.

\subsection{Size-of-Sample effect: Region Masses}
\label{sect:sample}

Large samples of star-forming regions are more likely to contain rare
massive members than are small samples. This should be true for any
subsample of a collection too. If the mass distribution function of
star-forming regions is $dN/dM\propto M^{-2}$, which is usually the
case, then the mass of the most-massive region should increase linearly
with the number of regions. Previous discussions in this paper
illustrated numerous examples of this size-of-sample effect. Figures
\ref{fig-radnmaxmass} and \ref{fig-radnmaxmass2} show two
results related to the size of sample effect: a trend in the maximum mass of a star-forming
region in a radial interval versus the surface density of regions (Fig.
\ref{fig-radnmaxmass}) and a trend versus the total number of regions (Fig.
\ref{fig-radnmaxmass2}) in that interval. In both cases, there is a positive
correlation such that the most massive regions are also where the
surface density and total count are highest, which tends to be the inner
part of a galaxy.  The clearest correlations have a slope of about
unity. This figure illustrates an effect where the star-forming regions
in the outer part of a galaxy tend to be lower mass than those in the
inner part, not because of a physical change with radius, but because
the total count of regions in the outer part is smaller.

We are also interested in whether the size-of-sample effect applies
when comparing one galaxy to another. Figure \ref{fig-nmx} shows the
maximum region mass plotted as a function of the total number of
regions extrapolated down to a mass of $10^3\;M_\odot$, which is the
lowest observable minimum mass for all of our galaxies. If the
observable minimum $M_{\min}$ for a particular galaxy is higher than
this, then the extrapolated count is taken to equal the count higher
than $M_{min}$ multiplied by the ratio $M_{min}/10^3\;M_\odot$.  This
would be the extrapolated count in the case when the mass function is
$\propto M^{-2}$.  The figure shows a nice correlation with a slope of
unity, as expected for the size-of-sample effect. Thus galaxies with
more massive star-forming regions are the ones with more total regions
(e.g. Billett, Hunter, \& Elmegreen, 2002; Whitmore 2003). The galaxies
have these massive regions only because the star-formation process is
able to sample further into the tail of the region mass distribution
function.  This result implies that star formation operates with
universal physical processes that are stochastic in nature. Of course,
the specific region masses are determined by these processes, but the
mixture of possibilities is so large that the mass distribution
function is essentially random when it is observed as an ensemble.

The one exception in our study seems to be the super star cluster in
NGC 2366, which is relatively unaccompanied by lower mass clusters (\S \ref{sect:2366}).
This giant cluster
could either have been formed by a completely different process, one
that does not give the same $M^{-2}$ mass distribution function as the
main process, or it could have grown to its current mass by coagulation
of lower-mass clusters. Models of cluster coalescence are in Elmegreen
et al. (2000) and Fellhauer \& Kroupa (2002).

\section{Summary}

With the small size of dwarf galaxies comes low surface brightnesses
and low densities. Yet despite these apparent impediments to star
formation, we see dwarf galaxies actively forming stars, sometimes in
dramatic starbursts. We also see star formation in the far outer parts
of dwarf galaxies, where the conditions are even more extreme. Here we
identified star-forming regions in nearby dwarf galaxies on UV images
in order to study their properties as a function of galactocentric
radius and from galaxy to galaxy.

We measured the ages and masses of the more prominent star-forming 
regions in 11 nearby dIm galaxies. We were able to define 1623 useable 
regions on \textit{GALEX} $NUV$ images. We see star forming regions 
that run the gambit from tens to millions of solar masses, and we were 
able to find regions as young as our lower limit of several
million years to our completeness limit of 500 Myrs.
Within our sample of galaxies we see stark differences 
from one to the next. For example, our closest galaxy has at least 713 
star-forming regions, while our next closest galaxy has only 5. Still, 
for all of our galaxies, the maximum region mass scales linearly with 
the total number of regions, and within each galaxy, the maximum mass 
in a radial annulus scales linearly with the local number in that 
annulus. Both trends indicate that star formation mass is independent 
of local environment and even the surrounding galaxy mass, but is given 
by random sampling from a universal mass function that is of the usual 
form $dN/dM\propto M^{-2}$. 

We find that our galaxies fall into three categories describing cluster 
formation activity over the past 500 Myrs: star formation with gaps, 
nearly constant star formation, and little-to-no star formation. The 
cases with gaps include some with occasional spikes in the star 
formation rate, and others with a cessation of star formation in the 
last 10 to 25 Myr. 

There is no consistent, measurable radial dependence for ages and
masses of star-forming regions. The exceptions, DDO 50 and NGC 2366, show a
trend towards lower masses at greater radii, but even there we see
massive regions on the edges of the optical galaxies.
Generally there is no indication that dwarf Irregular galaxies are less
capable of producing massive star-forming regions in their outer disks
than in their inner disks. Still, the regions in the outer disk often
have no discernable H$\alpha$ emission. This is presumably the result
of the low density there, which makes H$\alpha$ in the vicinity of
massive stars very difficult to observe.

These little galaxies have in the past presented many quandaries as to
how something so small and sparse can form stars, and now it looks like
they are forming stars in areas even more extreme than was previously
believed possible from other surveys. This is consistent
with a picture in which localized processes form stars in most regions
of dwarf galaxies. At locations where the average gas surface density
is much smaller ($\times10$) than the fiducial critical value, we still
see star formation. The detailed reasons for this are unknown. There
are no spiral distributions for this star formation as there are in the
outer parts of large spiral galaxies, so we cannot identify spiral wave
compression as a primary cloud formation mechanism.  It could be that
all of the outer galaxy star formation and much of what happens in the
inner region too, is triggered locally, by stray supernova, for
example, or by the turbulence that supernovae and other energy sources
drive. It could also be that large-scale gravitational instabilities
operate everywhere, even in sub-threshold gas, because of an easy
transfer of angular momentum away from growing density perturbations.
In that case, the collapse rate could be smaller than the local
dynamical rate, more like the angular momentum loss rate, but the final
result for the scale and properties of star formation could be the
same. High resolution observations of the gas in these regions should
clarify this picture.

The star forming regions in our survey also offer some insight into the
evolution of measureable mass with age. The distribution of regions on
a $\log M $ versus $\log t$ diagram is somewhat uniform over age $t$
for equal intervals of mass $M$. This uniformity suggests that the
regions are being removed from our view systematically with age. This
loss is consistent with a model in which each individual region mass is
decreasing as $1/t$. It is also consistent with another model in which
each region mass is constant until that region is suddenly disrupted,
and the disruption probability is proportional to $1/t$. Both
possibilities could happen simultaneously too. We cannot say anything
about the regions that have been disrupted, but we note from a trend in
the age dependence of surface brightness that measured masses may in
fact be decreasing as $1/t$.  This result is illustrated most clearly in
a diagram of fading-corrected absolute magnitude versus the log of the
age. A correlation in this diagram has a slope consistent with
$M\propto1/t$. The implication is that the outer parts of star-forming
regions are getting lost from our view as they slip below the surface
brightness limit of the survey. This is a loss of mass over and above
stellar evolutionary effects (supernovae and winds) and in addition to
evaporation.  We are investigating all of these mass loss processes in
more detail in a second paper.

\acknowledgments

We are grateful to Philip Massey for help with calibration of the Local Group Survey Project's M33 images,
which were used for $UBV$ photometry of the test object NGC 604.
We acknowledge a careful reading and useful comments from an anonymous referee.
Funding for this
research was provided to DAH, NM, and BGE by NASA-GALEX grant
NNX07AJ36G and by cost-sharing from Lowell Observatory. LZ participated
in the 2007 Research Experience for Undergraduates program at Northern
Arizona University (NAU). We appreciate Kathy Eastwood's efforts in
organizing that program and the National Science Foundation for funding
it through grant  AST-0453611 to NAU.

Facilities: \facility{{\it GALEX}}, \facility{Lowell Observatory}

\appendix

\section{Individual galaxies} \label{app-indiv}

The UV-bright knots identified in each galaxy are outlined in Figure \ref{fig-images}.
Here we discuss the properties of the ensemble of regions in each galaxy.

\subsection{NGC 6822}

NGC 6822 is the closest galaxy in our survey (0.5 Mpc). Because of
this, we were able to identify 800 individual star-forming regions, 713
of which made it into our final data set. With this galaxy we were able
to trace the star forming regions to much lower luminosity levels than
in any other galaxy. A problem arises because of the galaxy's large
angular size (about 15\arcmin$\times$15\arcmin), which led to field
star contamination at this low galactic latitude.

NGC 6822 has star-forming regions ranging in mass from 15 $M_{\sun}$ to
$3\times 10^5 M_{\sun}$, and ranging in age from several Myrs to well
beyond the 500 Myrs that is emphasized in the UV.
The lower mass
limit of a star-forming region that should still be visible after 500
Myrs is $5\times 10^3 M_{\sun}$ in this galaxy.
Figure \ref{fig-numvsage} suggests
that star formation has occurred in two or three major
episodes. These episodes appear as peaks in the number of all clusters
detected and in the number of clusters detected with masses larger than
our adopted lower mass limit.
They also appear
as peaks in the mass of the most massive cluster as a function of age.
The episodes occur for logarithmic ages between the present and 7.5, between
8.12 and 8.78, and after 9.28.  We refer to these as the 7
Myr, 100 Myr, and 1 Gyr bursts.

In the most recent star formation event at 7 Myr, there are 3
clusters more massive than $5\times10^3\;M_\odot$. The most massive
cluster in this event is $2\times10^4\;M_\odot$. In the event at 100
Myr, there are 7 regions of mass $\ge 5\times 10^3 M_{\sun}$, with the
most massive of these having a mass of $7\times 10^4 M_{\sun}$. In the
older burst at 1Gyr, which may not be complete in time and is certainly
not complete to our adopted lower mass limit,
we see 54 regions of mass $\ge 5\times 10^3
M_{\sun}$, with the most massive having a mass of $3\times 10^5
M_{\sun}$. The oldest age burst is old enough that all of the clusters
below $5\times 10^3 M_{\sun}$ have faded, and we see this in the lower
mass limit of the regions we identified.

We can estimate the apparent relative formation rates of star-forming
regions from the counts of regions on the time axis.  For the 7 and 100
Myr bursts, the apparent region formation rates with
$M\ge5\times10^3\;M_\odot$, obtained from the counts per unit time
interval, are 0.1 and 0.01 massive regions per Myr, respectively. These
are the apparent and not the real formation rates because many clusters
have been lost from their original mass intervals through disruption
and evaporation.

For a mass function of the form $dN/dM\propto M^{-2}$, the total number
of initial regions should be equal to the ratio of the maximum region
mass to the minimum region mass. This is one aspect of the
size-of-sample effect. If we consider a minimum detectable mass to be
$5\times10^3\;M_\odot$, then the number of regions above this mass
should be proportional to the maximum region mass if the disruption
probability is independent of mass (e.g., as observed by Chandar et al.
2009). From the pure counts of clusters above $5\times10^3\;M_\odot$,
the proportion in the two most recent bursts is 1:2.3, and from the
maximum masses, the proportion is 1:3.5. These proportions are close
enough to each other to agree with the predictions of the size of
sample effect, considering the small numbers of massive clusters.
(fractional errors in the counts are 60\% and 38\%, respectively, from
the inverse square roots of the number of massive regions).

Fall et al. (2005) consider the mass-age diagram for the Antenna galaxy
and note that generally the number of clusters above a certain mass in
equal log intervals of time is about constant with time, which means
that the number of clusters remaining per unit age interval decreases
as $1/{\rm age}$. In our sequence of ages, the counts per unit time,
0.1 and 0.01 regions Myr$^{-1}$ are indeed roughly proportional to the
inverse of their ages (7 and 100 Myr).  This implies that clusters of
all masses are becoming increasing lost over time. Such loss could be
from a combination of the two processes as discussed in \S
\ref{sect:destroy}, and from the surface brightness effects discussed
in \S \ref{sb}.

From $\log t = 7.9$ to $\log t = 8.1$ and from $\log t=8.8$ to $\log
t=9.3$, periods of 46 Myr and 1300 Myr, there are lulls in the star
formation rate. In the first interval, no regions of $M\ge 1\times
10^3\; M_{\sun}$ form and the maximum mass drops down to
$1\times10^3\;M_\odot$. In the second interval, no regions were formed
as massive as the periods on either side, but the maximum mass is
$1\times10^4\;M_\odot$. In the bursts before and after these lulls, the
maximum cluster masses were higher by factors of $\sim7-70$ and the total
numbers of clusters remaining were higher by factors of $\sim1-10$.
Thus we are seeing oscillations in the cluster formation rate by a
factor of order $\sim10$ in both the cluster counts and the maximum cluster
mass. This type of variation is consistent with Lee et al.'s (2007)
prediction.

The time intervals of the peaks and lulls decrease toward the present
time. This appears to be the result of star formation with a clustered
or hierarchical structure in time (for models of such star formation,
see Elmegreen \& Scalo 2006). Cluster bursts of short duration that
occurred long ago would appear blended in time because our time
resolution cannot resolve the small subbursts, while cluster bursts of
the same intrinsic width that occurred recently can still be resolved.
The time resolution increases in proportion to a power of the cluster
age because the cluster colors evolve with a power law time dependence.

Figures \ref{fig-plots} and \ref{fig-numvsage} suggest that the recent
formation of massive regions in NGC 6822 is more prominent in the outer
part of the disk, beyond $R_{H\alpha}$. The left-hand part of Figure
\ref{fig-numvsage} also suggests that most of the young regions,
regardless of mass, are outside $R_{H\alpha}$. In fact 468 of this
galaxy's 713 star-forming regions fall outside of \rha, which is far
more than we see in any other galaxy in our survey. This is consistent
with the distribution of blue stars (Komiyama \et\ 2003).

\subsection{IC 1613}

IC 1613 is our third nearest galaxy at 0.7 Mpc, and second most
populous galaxy with 342 useable star-forming regions. Because it is
nearby, we were able to trace the star-forming regions to masses as low as
$\sim30\;M_{\odot}$.
The minimum mass for which we should
see all currently definable star-forming regions from the last 500 Myrs
is approximately $3 \times 10^3 M_{\sun}$, as indicated by the dashed
horizontal line in Figure \ref{fig-plots}.
The maximum mass formed is of order $3 \times 10^4 M_{\sun}$.
In IC 1613 there has been more-or-less continuous star formation over
the last 1 Gyr.
The apparent formation rates of massive star-forming regions
($>3\times10^3\;M_\odot$ for IC 1613) from log age of 7.0 to 8.5
is 0.03 regions Myr$^{-1}$.

IC 1613 has an \ha\ profile that ends before the UV star-forming
regions end. The \ha\ only extends to a radius of 7.5\arcmin\ from the
center of the galaxy, and we see star-forming regions in the UV out to
a radius of 9.3\arcmin.  There are far more massive regions inside
$R_{H\alpha}$ than outside. The clusters beyond \rha\ tend
to be lower mass with a few notable exceptions among the oldest identified
clusters.

\subsection{WLM}

Wolf-Lundmark-Melotte (WLM) was the third most populous galaxy in our
survey with 191 regions marked, 165 of which made it into our final
data set. Its distance was typical at 1.0 Mpc. This galaxy is distinct
in having the oldest average star-forming regions in all our sample.
The cluster formation rate appears to have been relatively constant
until it stopped about 22 Myrs ago.
The minimum detectable mass at 500 Myrs is about $1\times10^3\;M_\odot$.

This galaxy has a large fraction of star-forming regions beyond \rha:
31 out of the total 165. UV regions are seen to a radius of 7.4\arcmin,
while \ha\ drops out beyond 4.2\arcmin. In Figure \ref{fig-plots},
there appears to be a slight drop in the mass beyond $R_{H\alpha}$,
amounting to a factor of $\sim3$.

\subsection{DDO 50}

DDO 50 was the second most distant galaxy in our survey at 3.4 Mpc, but
we were still able to define 169 regions, 139 of which made it into the
final data set. Had this galaxy been located as close as WLM, from
the number of massive regions we see now and our experiments with
WLM images, we might expect to have
identified 520 regions instead of only 169.
Cluster formation in this galaxy appears to have been relatively constant
until about 9 Myrs ago. However, there is  a small enhancement in
the numbers of massive clusters 23-30 Myrs compared to the period of
time before that. The most massive cluster we identified is
$1\times10^5$ M\solar.

Because \ha\ extends nearly as far as star formation seen in the UV
(4.1\arcmin\ for \ha\ and 5.1\arcmin\ for the UV), we found only 10
regions beyond $R_{H\alpha}$. The outer regions sample the full
distribution of masses despite the small sample size, but they tend to
be a little lower in mass on average compared to those in the rest of the galaxy.

\subsection{DDO 75}

DDO 75, also known as Sextans A, is another populous galaxy (119
useable regions) at a moderate distance (1.3 Mpc). One of the unusual
characteristics of this galaxy is that its azimuthally-averaged
$V$-band and 4.5-micron surface
brightness profiles do not fall off at a constant exponential rate.
Instead, we see a flat profile in the inner 1.8\arcmin\ radius (0.7
kpc), followed by a normal exponential decay curve beyond that (Hunter
\& Elmegreen 2006). To investigate whether the break in the profile
represents a change in star formation characteristics, we compare the
regions interior and exterior to the break in Figure \ref{fig-d75}. We
plot mass versus age for the inner and outer disks in the top part of
the figure, the number of regions in radial intervals of 0.25 arcmin
versus radius in the bottom left, and the average $M_V$ versus radius
in the bottom right. In none of these plots do we see a significant
change in the star-formation characteristics at the surface brightness
break. There is a peak in the surface density of star-forming regions
at the break, but no obvious change in region properties.  This result
implies that the processes creating the break in the exponential
$V$-band surface brightness profile do not affect the masses or the
ages of the star-forming regions.  Most of the regions, and all of the
massive regions, formed inside $R_{H\alpha}$, so in Figure \ref{fig-plots}
we present a comparison of the inner half of the
star-forming regions to the outer half.

DDO 75 appears to have had a fairly constant level of star formation
for the last several hundred million years until about
9 Myrs ago. In the past 9 Myrs only one cluster has been formed.
Furthermore, clusters older than about 200 Myrs are missing
except for a few especially massive ones: clusters up to $4\times10^4$ M\solar\
in mass.

The apparent formation rate above the detectable mass limit of
$2\times10^3 M_\odot$ for ages between 9 Myr and 200 Myr was
$\sim0.08$ Myr$^{-1}$. The maximum cluster mass is $2\times10^4
M_{\sun}$. This steady low rate is consistent with the global view of
the galaxy from {\it HST} images and color-magnitude diagram fitting
(Dolphin \et\ 2003).

\subsection{NGC 2366}
\label{sect:2366}

NGC 2366 is among the more distant galaxies in our survey, at 3.2 Mpc,
but despite this we were able to define 66 regions, 58 of which were
useable. This galaxy also has $JHK$ data. NGC 2366 contains the most
massive star cluster in our combined sample, one with
$M\sim7\times10^6\;M_{\sun}$. A second cluster has a
mass of $2 \times 10^5 M_{\sun}$. These two regions are I
and II, respectively, in Drissen \et\ (2000). Drissen et al. split
cluster I into two subclusters, A and B, but this divide was not
discernible in our \textit{GALEX} images, most likely because IA is
still completely enveloped in dust and gas. Drissen \et\ find ages of
10 Myrs for cluster II, 2.5-5 Myr for cluster IB and an age less than 1
Myr for cluster IA. Our ages are significantly older for both regions,
40 Myrs for cluster II and 700 Myr for cluster I. Our older ages
may be due in part to our underestimating the reddening, which leads to
intrinsic colors that are too red. Another factor is the relative sizes
of the apertures we used. Our larger apertures encompassed the entire
complexes, including outlying cluster members. Billett {\et} (2002)
also looked at cluster I, or at least two smaller chunks of it. They
re-examined IB from Drissen {\et} and what they called cluster
1, both of which were given radii of 4.6 pc -- much smaller than the
200 pc radius used for the whole complex here. They also find
$M_{V}=-9.5$ for cluster IB and $M_{V}=-8.5$ for cluster 1, as compared
to $M_{V}=-14.4$ for the whole complex here.

The minimum mass for which we expect completeness up to 500 Myrs of age
is $1\times 10^4 M_{\sun}$, about the same as for DDO 50. The number of
clusters more massive than this is also about the same in these
galaxies (21 for NGC 2366 and 31 for DDO 50), and the most
massive cluster is extremely massive in both: $7\times10^6\;M_\odot$ in
NGC 2366 and $1\times10^5\;M_\odot$ in DDO 50.
A surprising
difference is that NGC 2366 has far fewer low mass clusters in the 10
Myr to 100 Myr age interval, 19 smaller than $10^4\;M_\odot$ in NGC
2366 compared to 93 of this mass in DDO 50.  The cluster mass function
is apparently flatter in NGC 2366, or more concentrated in its most
massive two clusters.  This difference could be the result of cluster
coalescence in NGC 2366, which would explain both the lack of low mass
clusters and the excessive mass of the largest cluster in that galaxy.
Cluster coalescence was shown to be reasonable for another
$10^6\;M_\odot$ cluster in NGC 6946 (Elmegreen, Efremov \& Larsen
2000).

This galaxy has only one $NUV$ region beyond
$R_{H\alpha}$. We present instead a comparison of the inner half of the
star-forming regions to the outer half in Figure \ref{fig-plots}. We
see several low mass, young star-forming regions in the outer half that
are absent in the inner galaxy, and in the very outer galaxy (past 4
kpc radius) we see no massive regions. NGC 2366 is a good example of a
galaxy with a gradient in the star formation region mass. Inside 4 kpc,
the $>10^4 M_{\sun}$ regions are uniformly distributed.

\subsection{DDO 70}

DDO 70 is a small galaxy (47 useable regions) at a moderate distance
(1.3 Mpc). It has had steady, slow star formation over the
last 200 Myrs, with peak masses of only $\sim8\times10^3 M_{\sun}$.
Like DDO 75, DDO 70 has formed only a few clusters over the past 10 Myrs
and only a few clusters before 200 Myrs ago.
There are no star-forming regions found beyond $R_{H\alpha}$, and so no
comparison can be made between the galaxy interior and exterior to
$R_{H\alpha}$. A plot of age versus mass is presented in Figure
\ref{fig-plots} with regions color-coded according to the inner 50\%
and the outer 50\% of regions.

\subsection{DDO 216}

DDO 216 is a very faint little galaxy. Despite its relatively low
distance (0.9 Mpc), we were able to identify only 34 regions, 25 of
which have made it into the final data set. This galaxy has a double
exponential $V$-band surface brightness profile: The profile interior
to a radius of 5.4\arcmin\ drops at one rate, and the profile exterior
to that radius falls off at a faster rate. However, there are no
UV-identified star-forming regions beyond a radius of 5.4\arcmin; our
furthest region was only 3.0\arcmin\ radius from the center of the
galaxy (deprojected distance).

DDO 216 has had very little star formation over the last 20 Myrs, but
before that there was a steady stream of modest star formation back to
at least 500 Myr. The lack of young star clusters accounts for the
small cutoff radius, $R_{H\alpha}$ (which is 1.3\arcmin), and
consequently most of the UV-detected star-forming regions are exterior
to \rha: 14 of 25 are beyond \rha. The only difference in the regions
inside and outside $R_{H\alpha}$ is that those outside extend to older
ages.
We see no difference in the minimum age or maximum mass of
clusters inside and outside of \rha, as shown in Figure
\ref{fig-plots}.

\subsection{DDO 210}

DDO 210 is a tiny little wisp of a galaxy. It is the second faintest
galaxy in our sample at an $M_V$ of $-10.9$, significantly fainter than
some of our super massive star clusters. Furthermore, this galaxy has
no HII regions. Because of its extreme faintness, and despite its
relative proximity (0.9 Mpc), we were able to find only 24 regions in
the $NUV$ image, and only 9 of those made it into the final data set.
Of these nine regions, one is less than 10 Myrs old, and 6
are 100 Myrs old or older. The lack of H$\alpha$ for the youngest
regions reinforces our suggestion elsewhere in this paper that
H$\alpha$ is too faint to see despite the likely presence of ionized
gas. Figure \ref{fig-plots} separates the regions into the inner 50\%
and outer 50\%.

\subsection{LGS3}

LGS 3 has only 7 definable regions, 5 of which were useable. This was
the second closest galaxy in our survey at a distance of 0.6 Mpc, so we
should be able to see any star formation that might have been there.
Despite these favorable conditions we still see nearly no star
formation over the last billion years. This galaxy has no measurable
HII regions.

\subsection{M81dwA}

For this tiny little galaxy we were able to find only three
star-forming regions, two of which were useable. M81dwA is the most
distant galaxy in our survey at 3.6 Mpc. The two regions we do see are
reasonably large, but between the two of them, they compose almost the
entire galaxy in the $NUV$. This is another galaxy without any
measurable HII regions. As we have only two data points, statistical
comparisons are not possible.

\clearpage

\begin{figure}
\epsscale{1.0}
\plotone{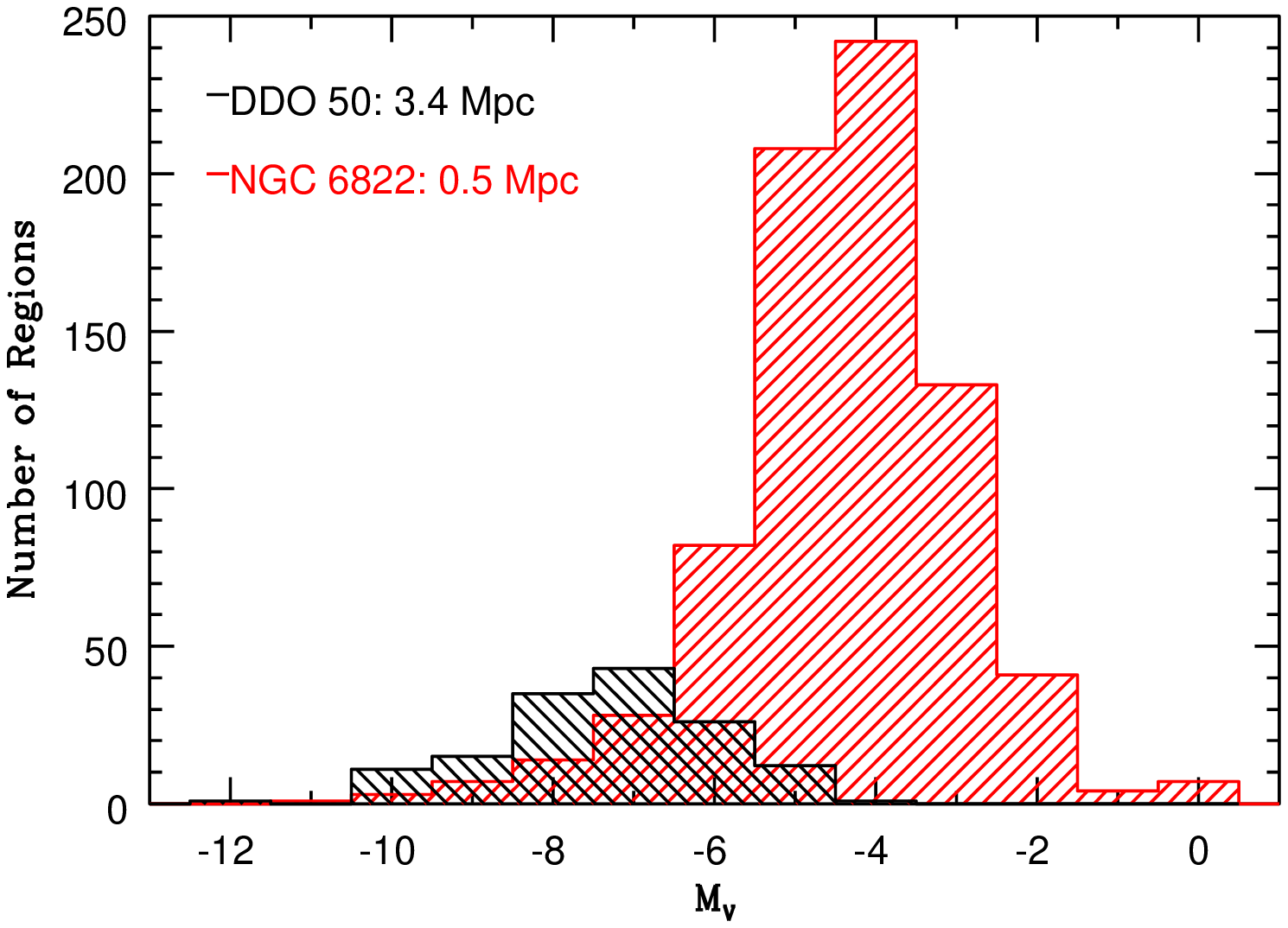}
\caption{Comparison of brightness distributions at different distances.
The number of regions as a function of $M_V$ of the regions is shown for DDO 50,
which is at a distance of 3.4 Mpc, and for NGC 6822, which is much closer at 0.5 Mpc.
The affects of distance and blending
on our detection limits is evident in the larger percentage of
fainter regions detected in NGC 6822.
\label{fig-disthist}}
\end{figure}

\clearpage

\begin{figure}
\epsscale{1.0}
\plotone{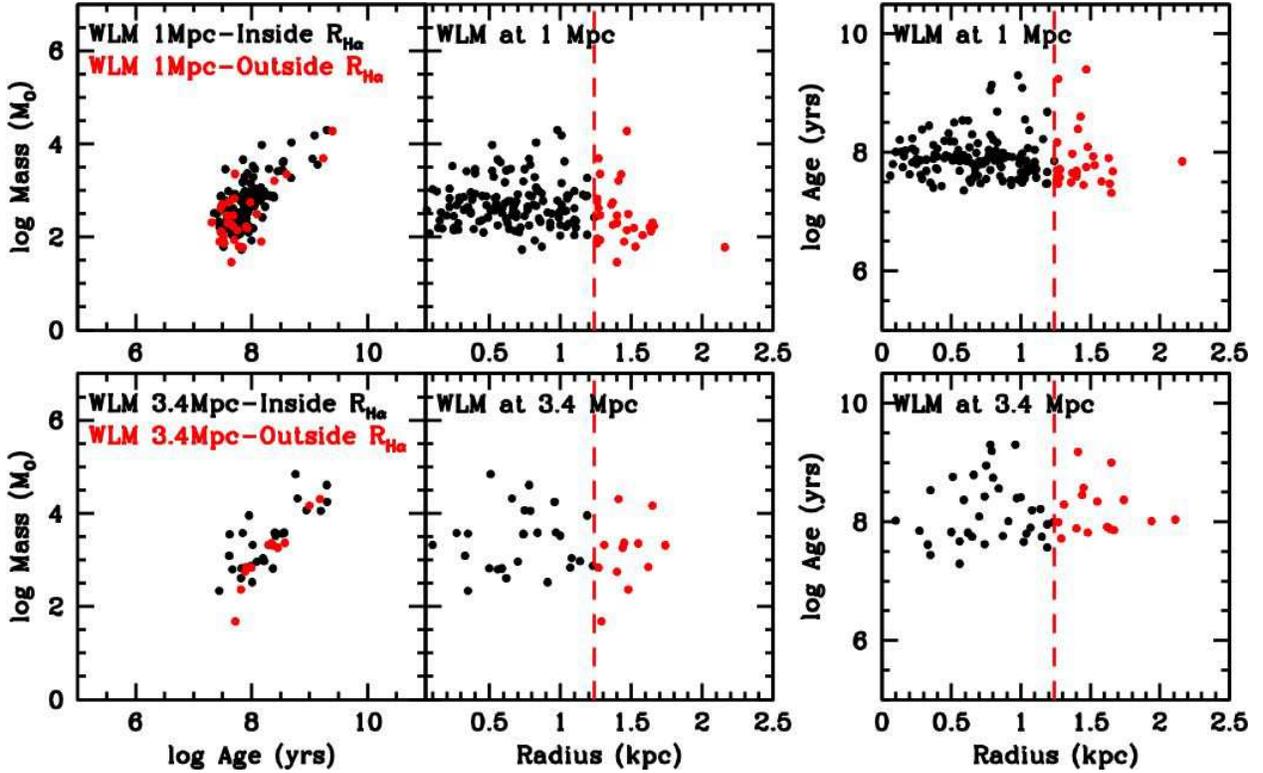}
\caption{Plots of mass vs.\
age, mass vs.\ galactocentric radius, and age vs.\ radius for WLM.
The nominal distance of WLM is 1 Mpc, and the panels labeled as
1 Mpc refer to regions identified and measured on the original
images. The panels labeled as 3.4 Mpc refer to regions
identified and measured on images that have been smoothed
to the resolution we would have if the galaxy were located
at 3.4 Mpc.
Regions outside the extent of \protect\ha\ are in {\it red}.
The uncertainties in the ages and masses have been left off in order to make the
plot less crowded.
At 3.4 Mpc, we see fewer regions with more mass and shallower radial gradients, as expected,
but general trends remain.
\label{fig-wlm}}
\end{figure}

\clearpage

\begin{figure}
\epsscale{0.9}
\plotone{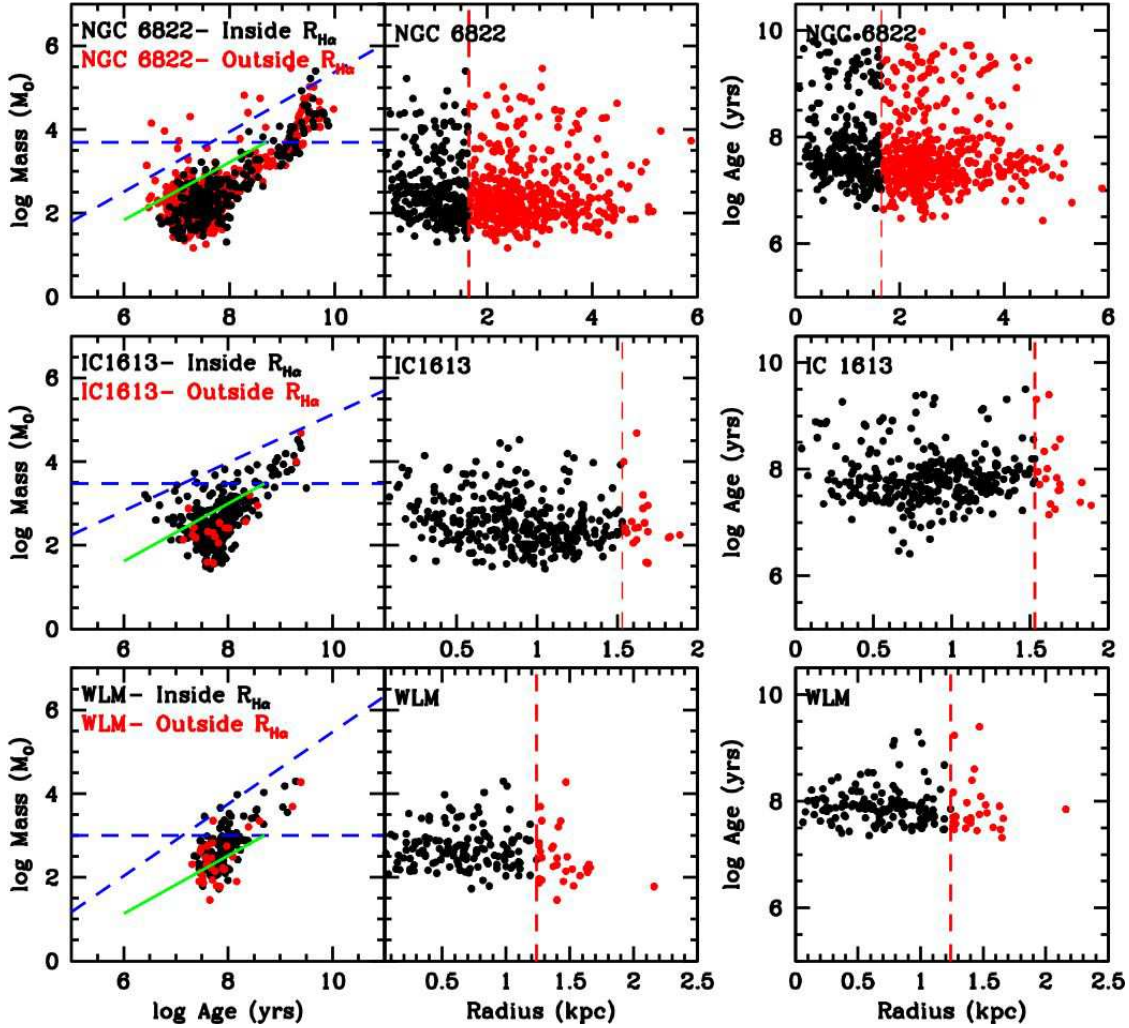}
\caption{Plots of mass vs.\
age, mass vs.\ galactocentric radius, and age vs.\ radius for the galaxies in our survey.
Regions outside the \protect\ha\ extent or in the outer half of all
regions are in {\it red}, as indicated by the labels.
The horizontal dashed lines in the
Mass-Age plots are mass limits for completeness to an age of 500 Myrs.
The slanted dashed lines are fits by eye to the upper envelopes of the
cluster distributions. The slanted solid line shows the slope for a
fading relationship in which the minimum observable mass scales as
$\log {\rm Mass}\propto 0.69\log {\rm Age}$. The vertical dashed lines
correspond to radii that separate the inner and outer parts, as defined
by the colored symbols.
\label{fig-plots}}
\end{figure}

\clearpage

\plotone{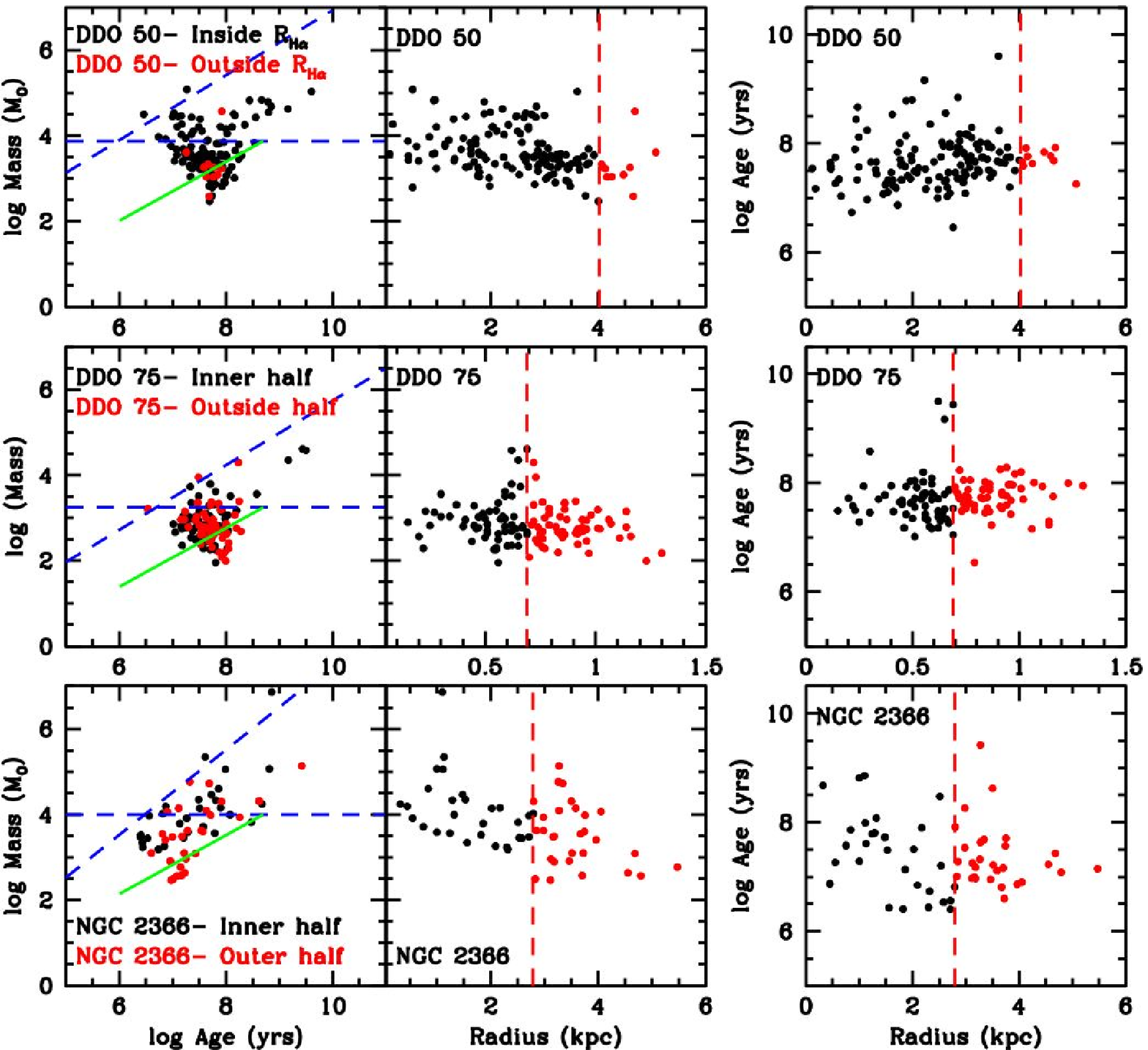}

Figure \ref{fig-plots} (continued)

\clearpage

\plotone{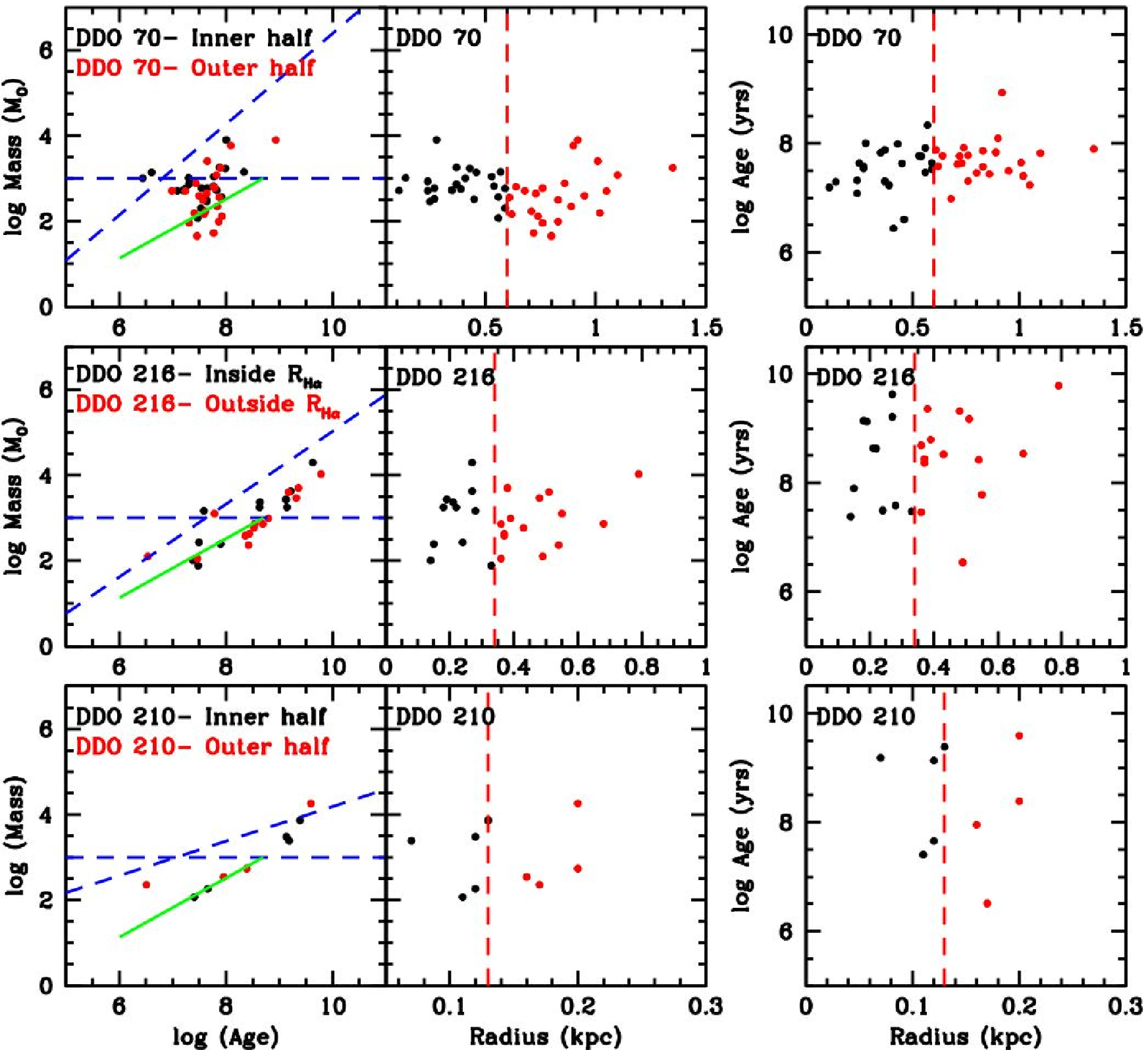}

Figure \ref{fig-plots} (continued)

\clearpage

\plotone{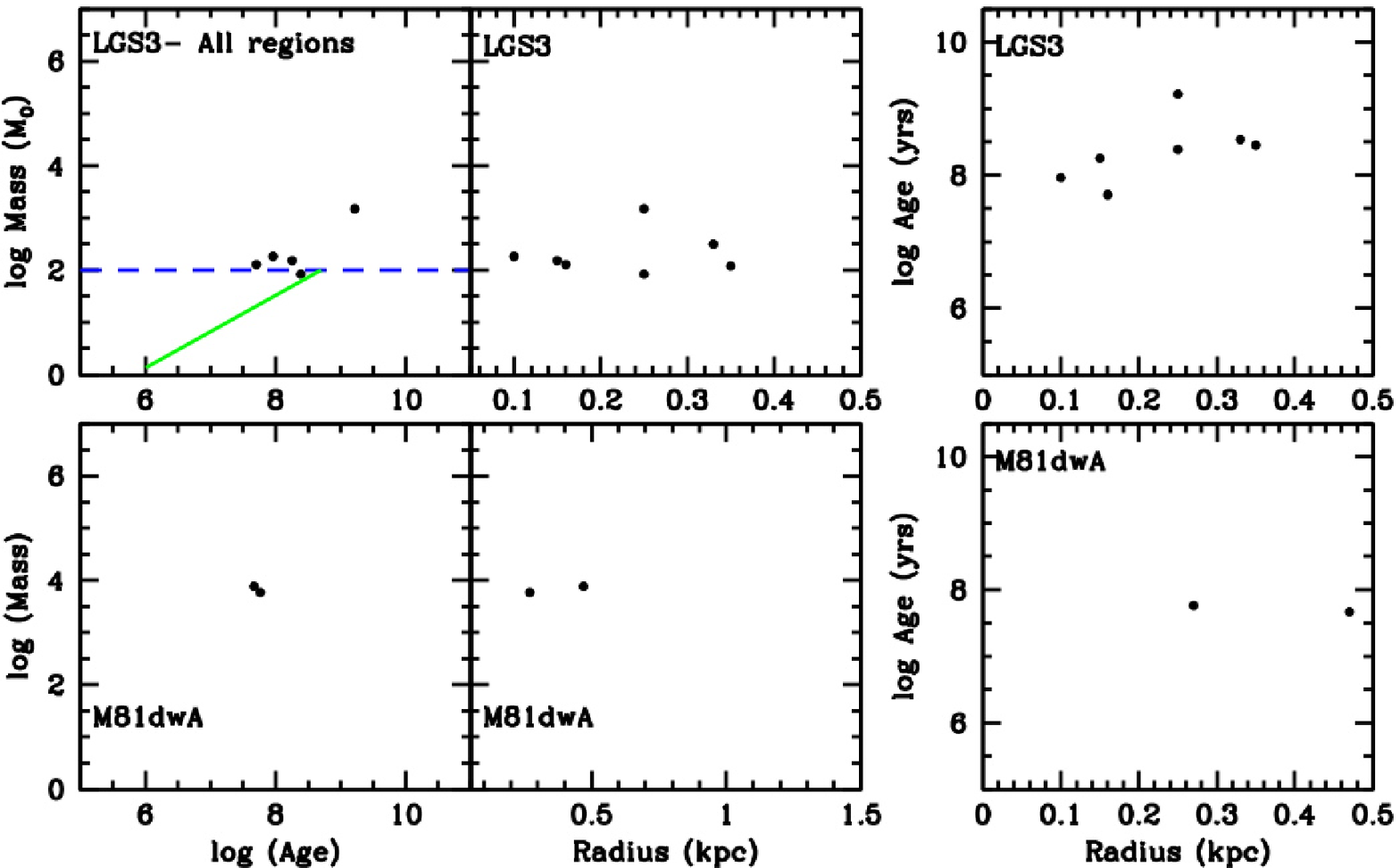}

Figure \ref{fig-plots} (continued)

\clearpage

\begin{figure}
\epsscale{0.95}
\plotone{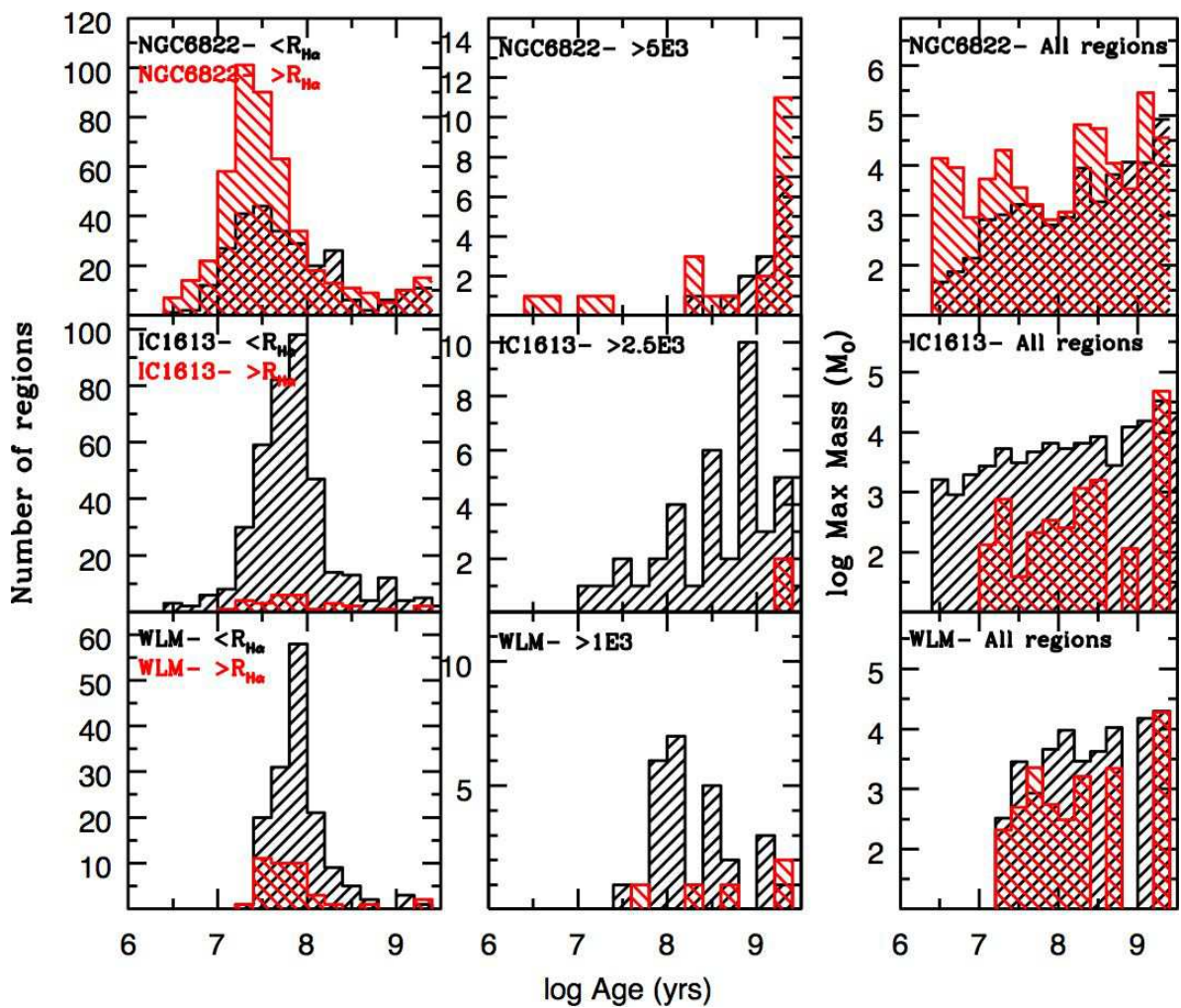}
\caption{Total number of
regions in our survey and the number larger than the minimum observable
mass (as indicated by each label) shown as a function of region age
with an age interval of $\Delta \log {\rm Age}=0.2$. On the right, the
maximum region mass in each age interval is shown. {\it Red} histograms are
for outer galaxy star formation, in the sense defined by Figure
\protect\ref{fig-plots}.
\label{fig-numvsage}}
\end{figure}

\clearpage

\epsscale{0.95}
\plotone{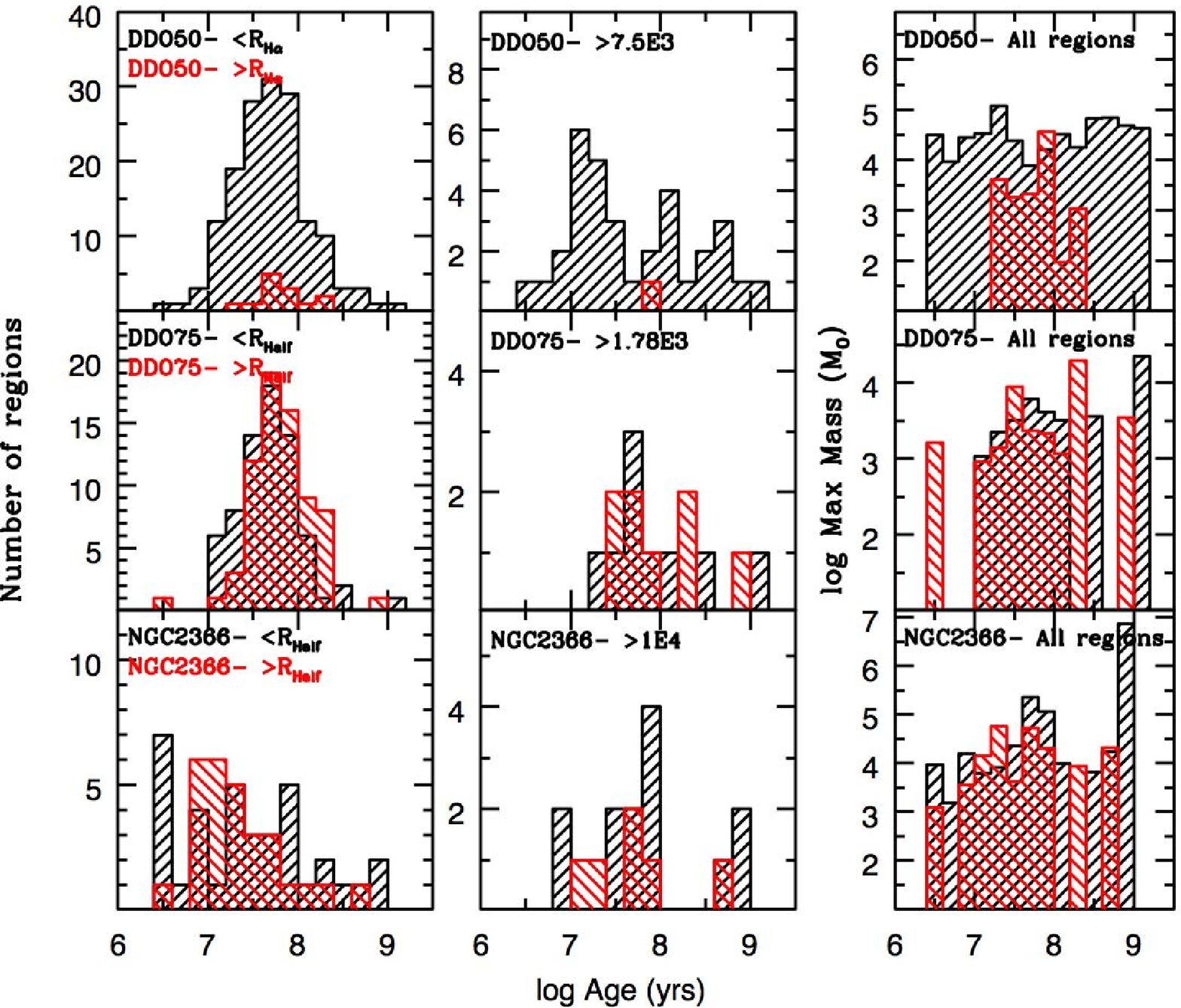}
Figure \ref{fig-numvsage}
(continued)

\clearpage

\epsscale{0.95}
\plotone{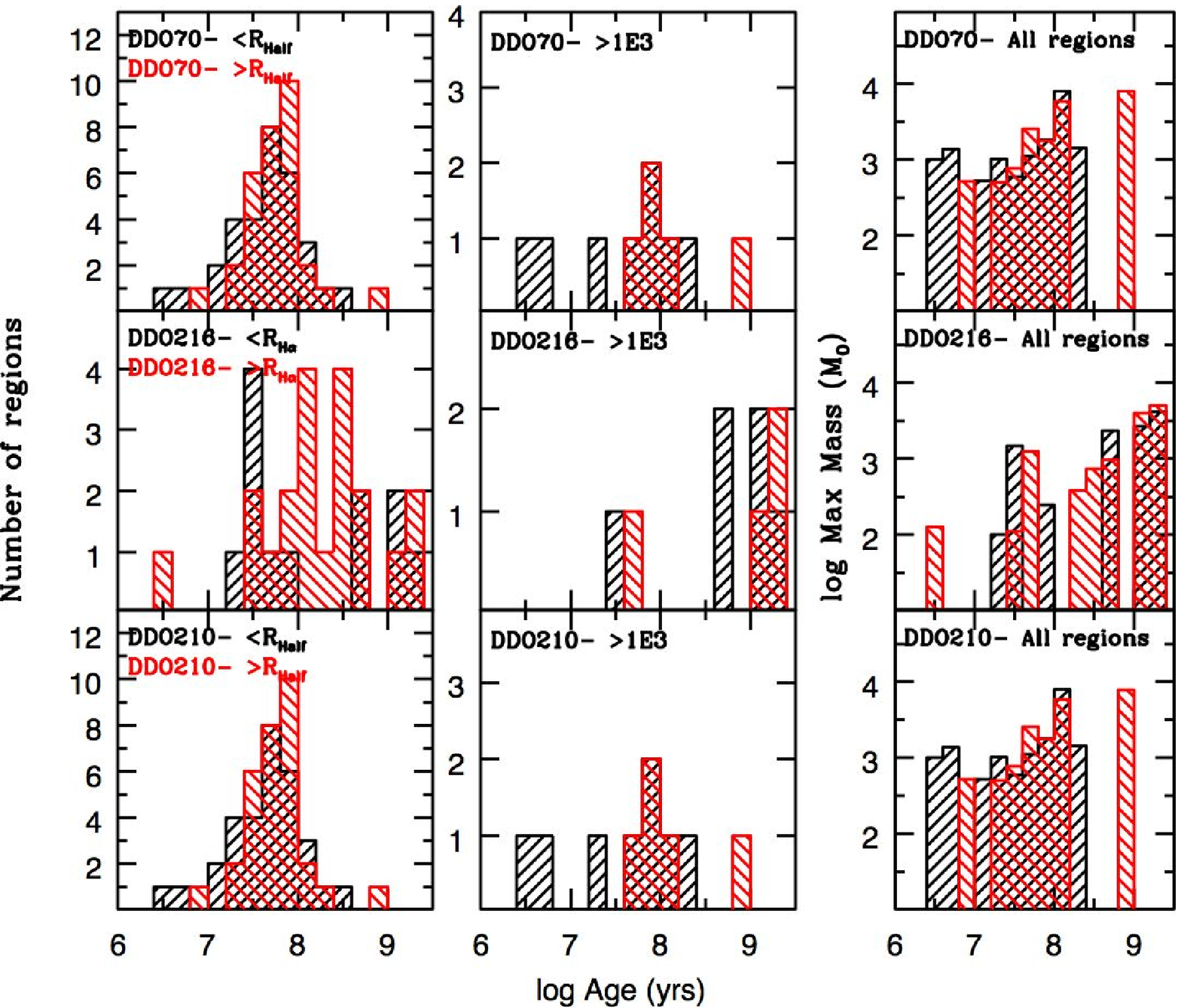}
Figure \ref{fig-numvsage}
(continued)

\clearpage

\begin{figure}
\epsscale{0.85}
\plotone{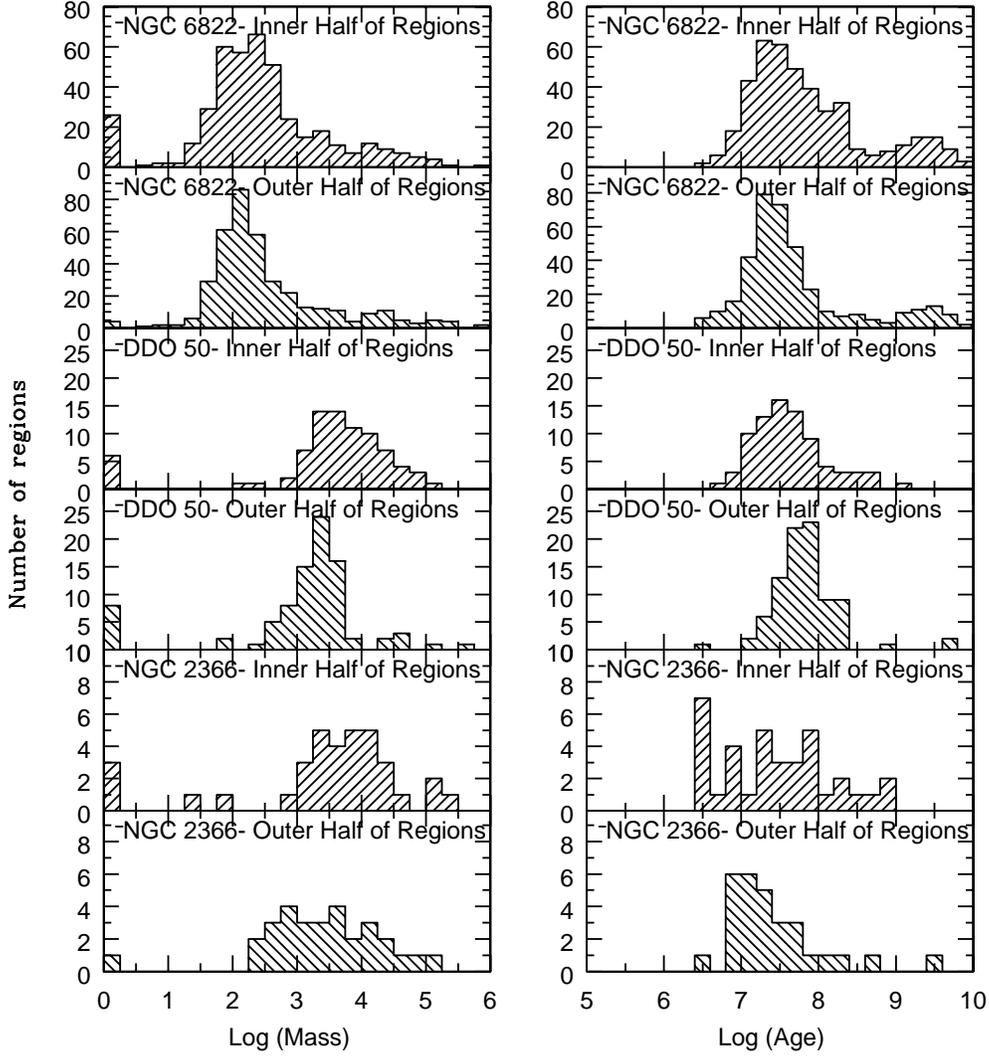}
\caption{
Mass ({\it left}) and age ({\it right}) distributions for the inner and
outer halves of star-forming regions in NGC 6822 ({\it top two panels
of figure}), DDO 50 ({\it middle two panels}), and NGC 2366 ({\it bottom
two panels}). The inner half of the regions is in the top panel for each
galaxy, and the outer half is in the bottom panel.
\label{fig-histd50n6822}}
\end{figure}

\clearpage

\begin{figure}
\epsscale{1.0}
\plotone{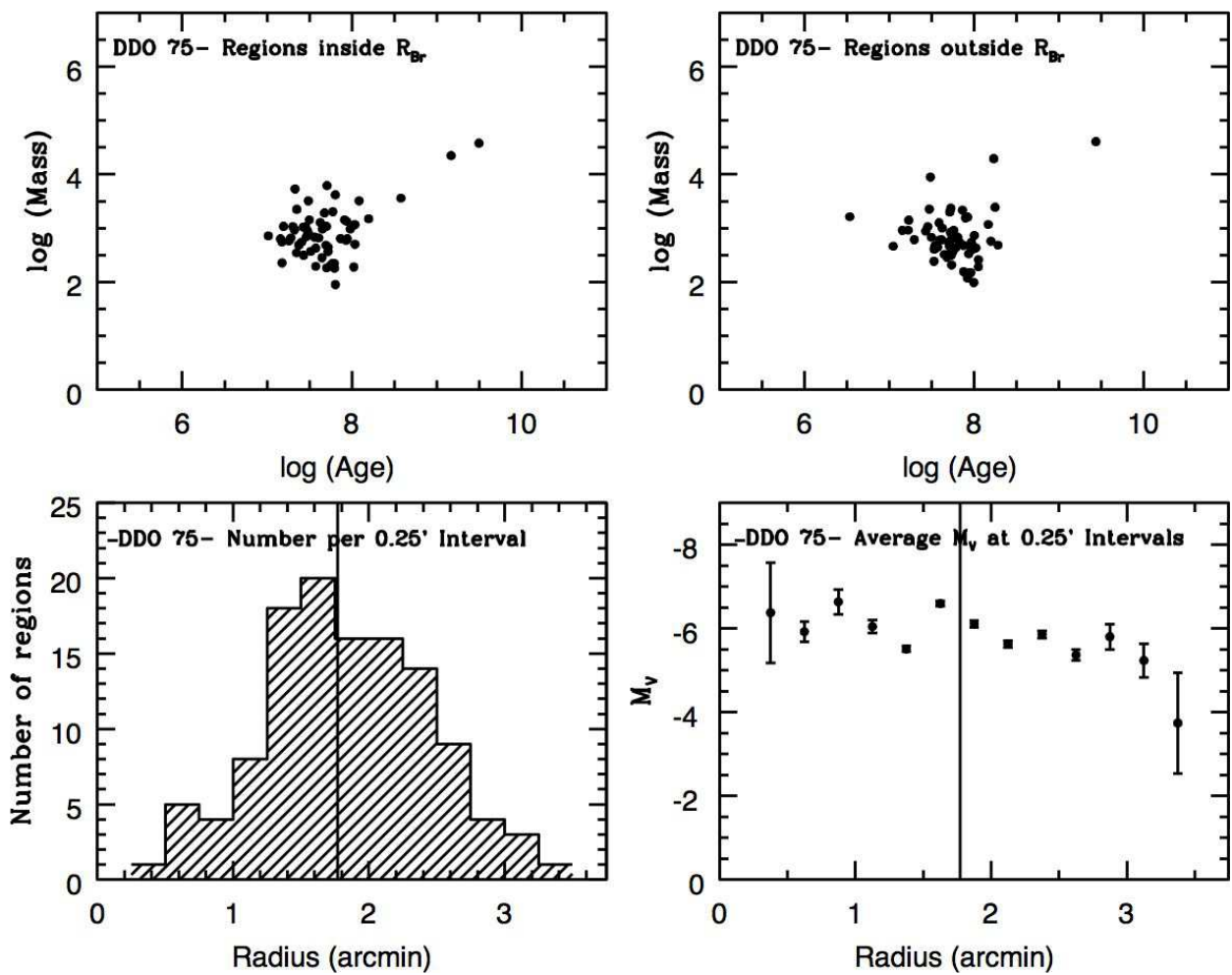}
\caption{Distributions of
star formation interior and exterior to the surface brightness break $R_{Br}$ in
DDO 75. In the {\it upper left} are plotted the masses and ages of regions
inside the break, and in the {\it upper right} are plotted the masses and
ages of regions outside the break. The {\it lower left} plot is the number of
star-forming regions per 0.25\arcmin, and the {\it lower right} plot is the
average $M_V$ for the star-forming regions in 0.25\arcmin\ bins.
The vertical line in the lower panels marks $R_{Br}$.
\label{fig-d75}}
\end{figure}

\clearpage

\begin{figure}
\epsscale{1.0}
\plotone{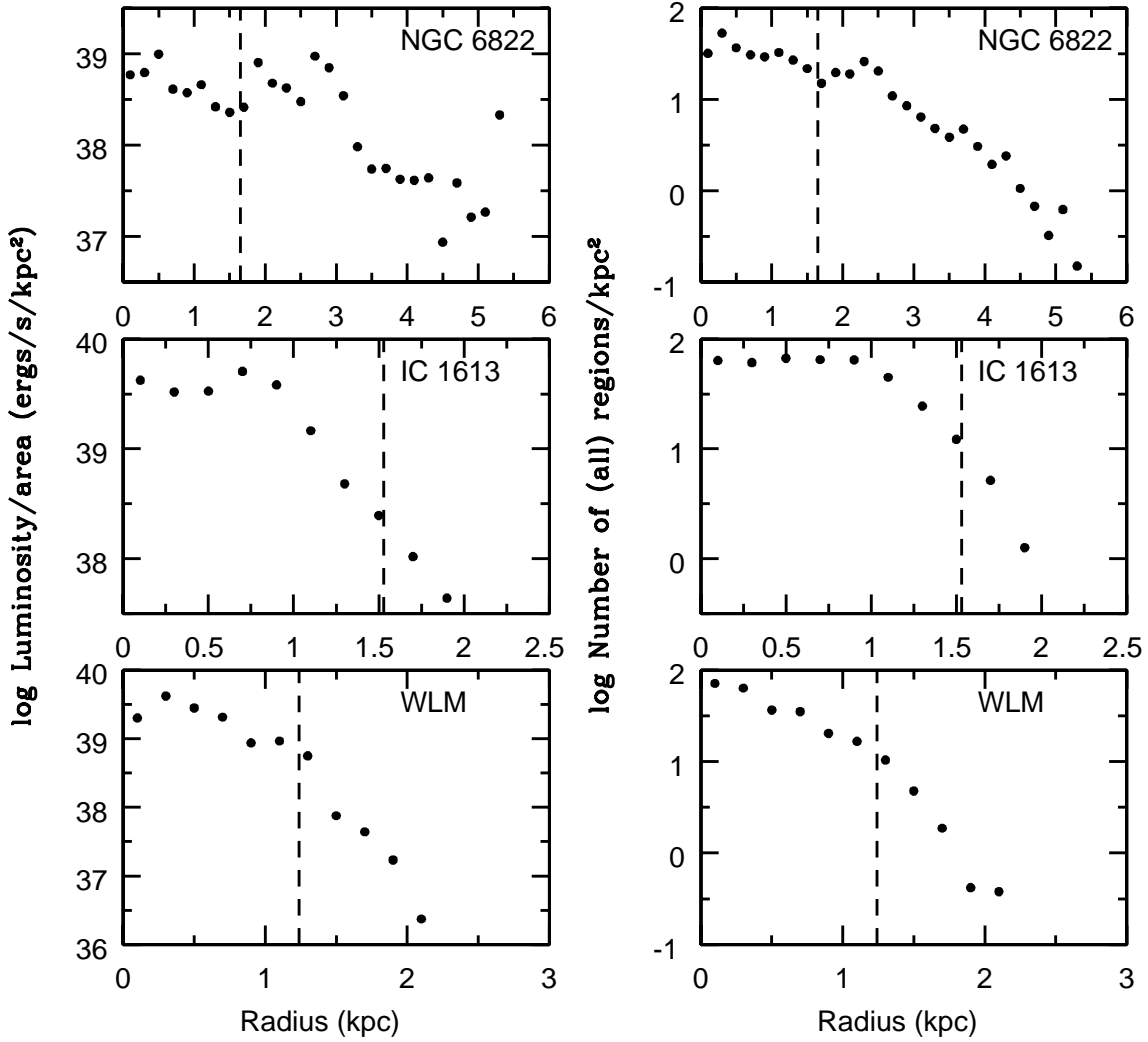}
\caption{Average $NUV$ surface brightness
and number density of all the identified regions shown as
functions of galactocentric radius for each galaxy with a large number
of regions. The vertical dashed lines correspond to radii that separate
the inner and outer parts, as defined in Figure \protect\ref{fig-plots}.
\label{fig-lumrad}}
\end{figure}

\clearpage

\plotone{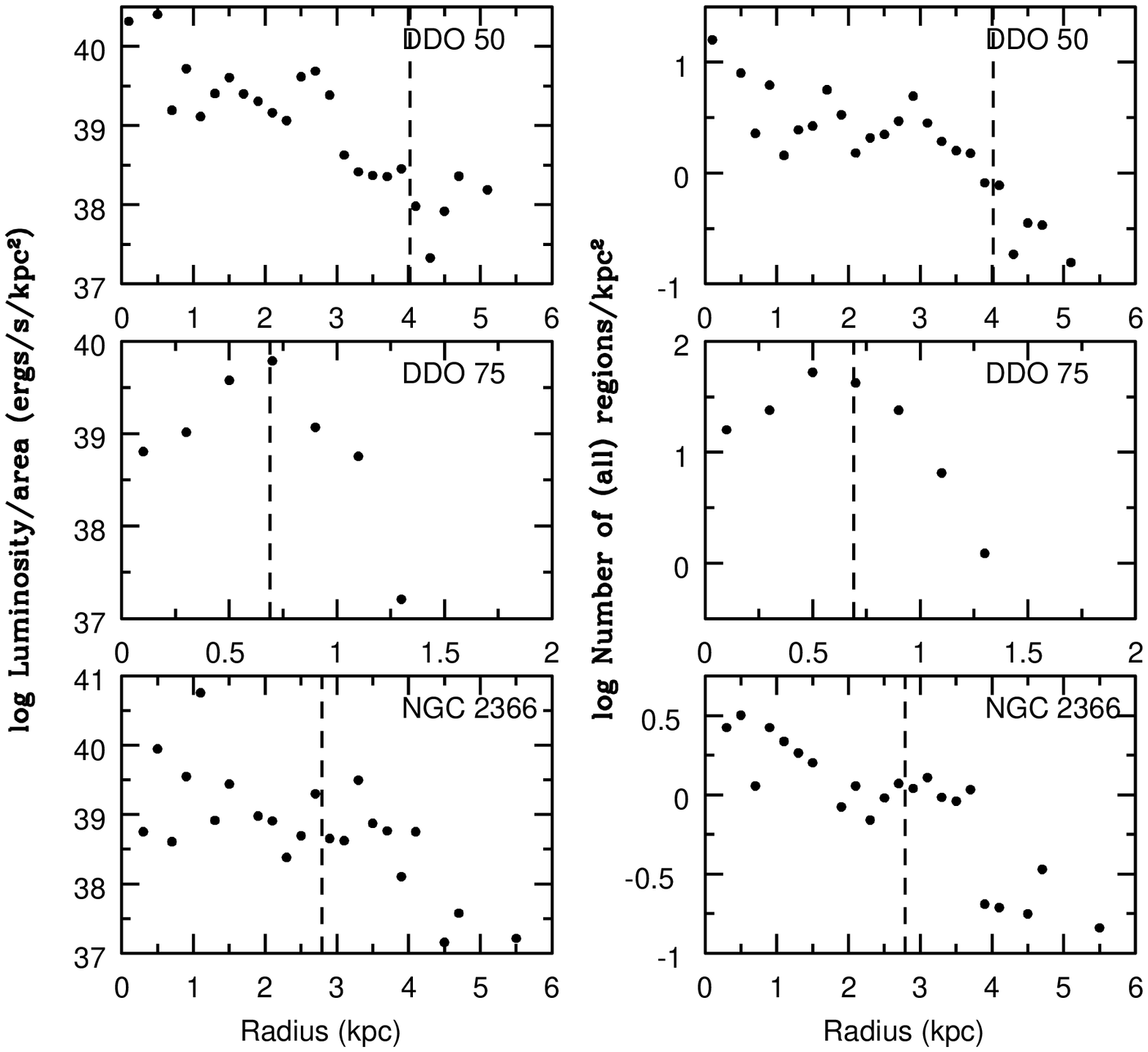} Figure \protect\ref{fig-lumrad} (continued)

\clearpage

\plotone{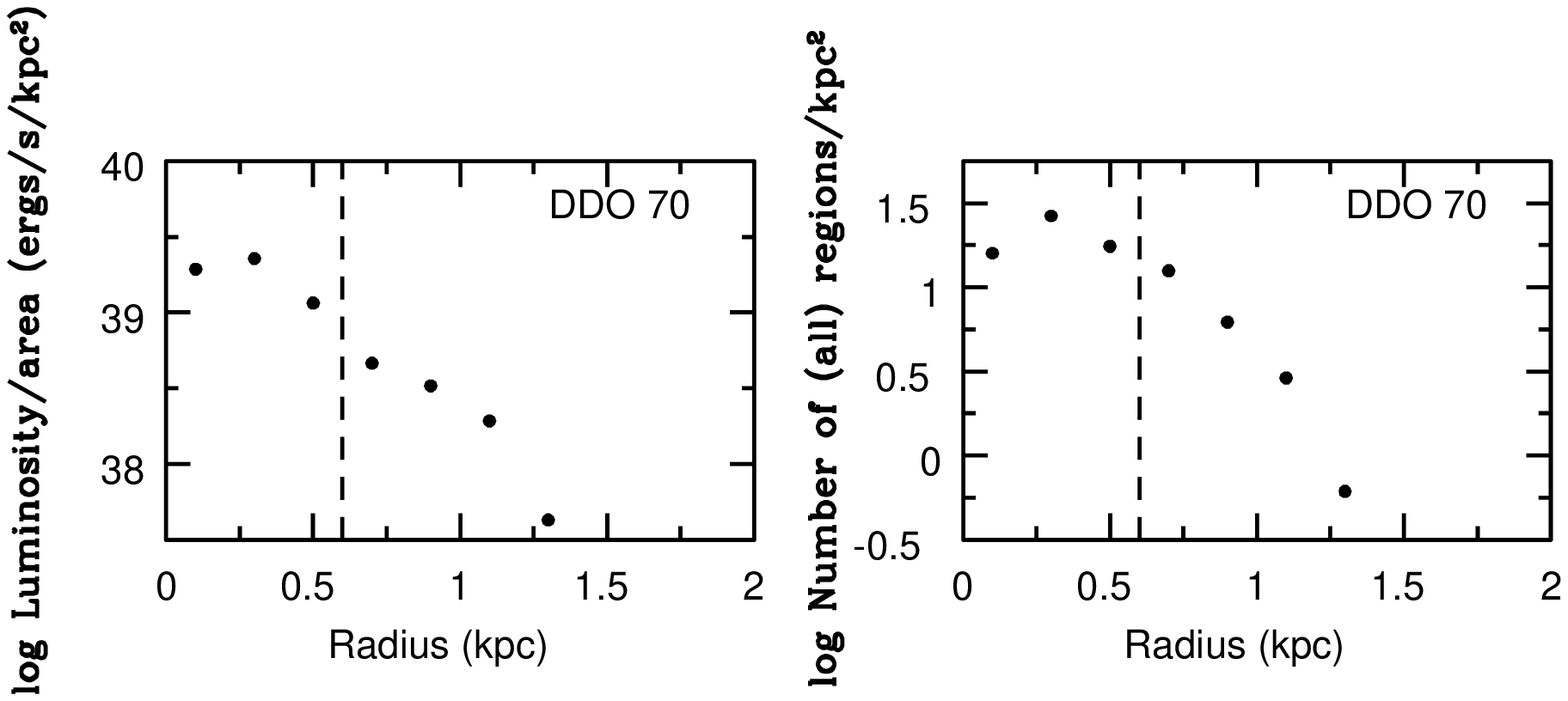} Figure \protect\ref{fig-lumrad} (continued)

\clearpage

\begin{figure}
\epsscale{1.0}
\plotone{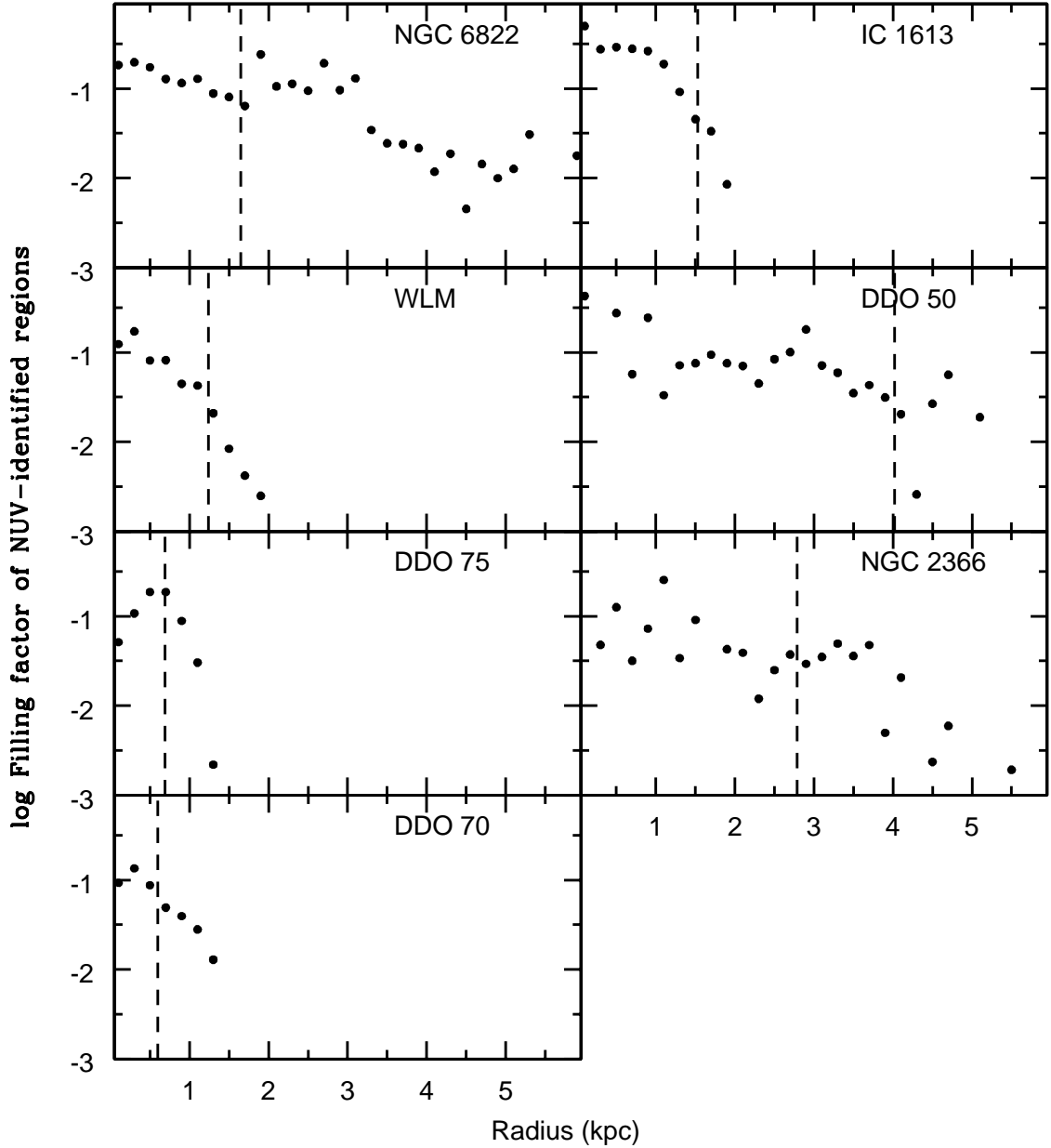}
\caption{Filling
factor of star-forming regions shown versus galactocentric radius.
The filling factor equals the sum of the region areas divided by the
area of the radial annulus. The trends here follow the radial trends in
luminosity density and counts because the UV region sizes are about
constant with radius.  The vertical line separates the inner and outer
regions as defined in Figure  \protect\ref{fig-plots}.
\label{fig-fillingfactor}}
\end{figure}

\clearpage

\begin{figure}
\epsscale{1.0}
\plotone{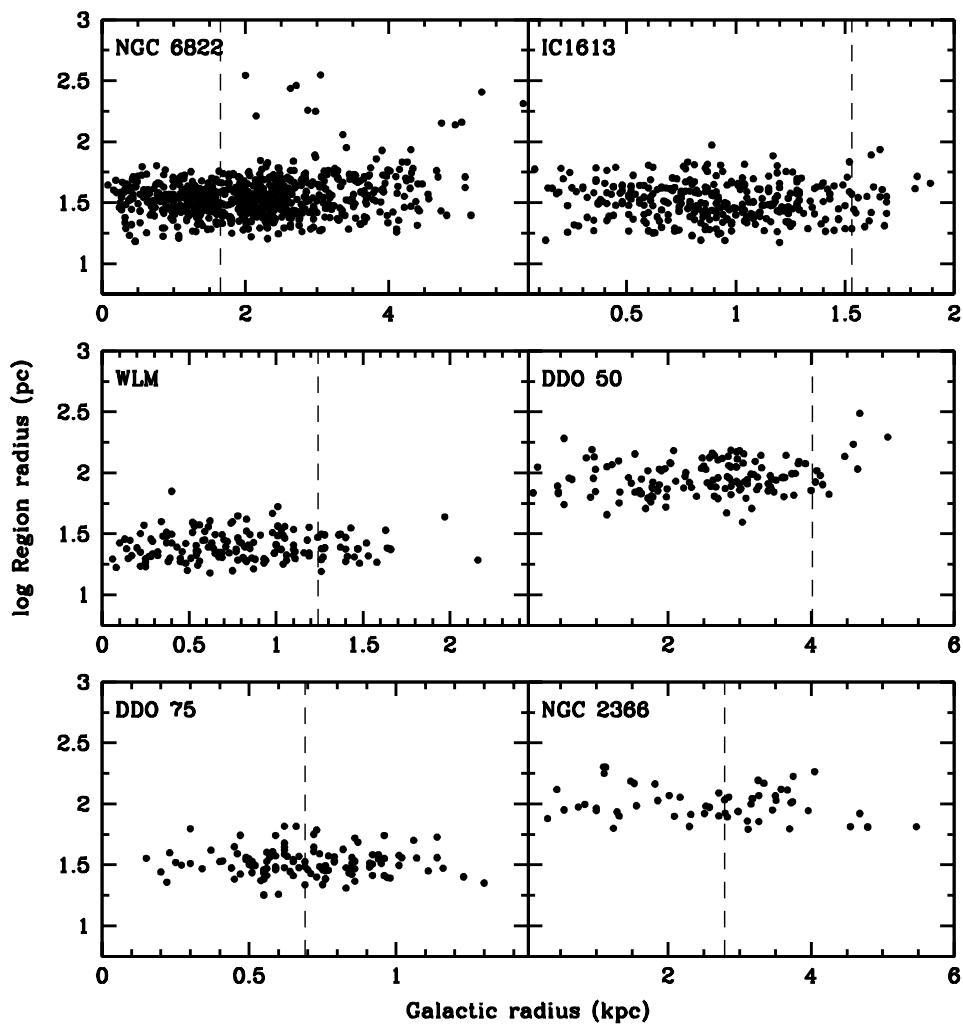}
\caption{Radius of each
region plotted versus the galactocentric radius. The region sizes
are estimates based on the polygons used to visually identify each
star-forming region in the $NUV$ {\it GALEX} images. They are not accurate
measures of region size. Still, they should reveal systematic
differences of a factor of two or more in the sizes for the inner and
outer parts of the galaxy (as delineated by the vertical lines), and no
such differences are observed.
The regions in DDO 50, NGC 2366, and M81dwA are significantly larger
compared to regions in other galaxies in our sample.
This is likely due at least in part to the larger distances of these galaxies,
and the difficulty in defining star-forming regions, especially faint,
small regions, to the same degree
of precision as was possible in closer objects.
\label{fig-radii}}
\end{figure}

\clearpage

\plotone{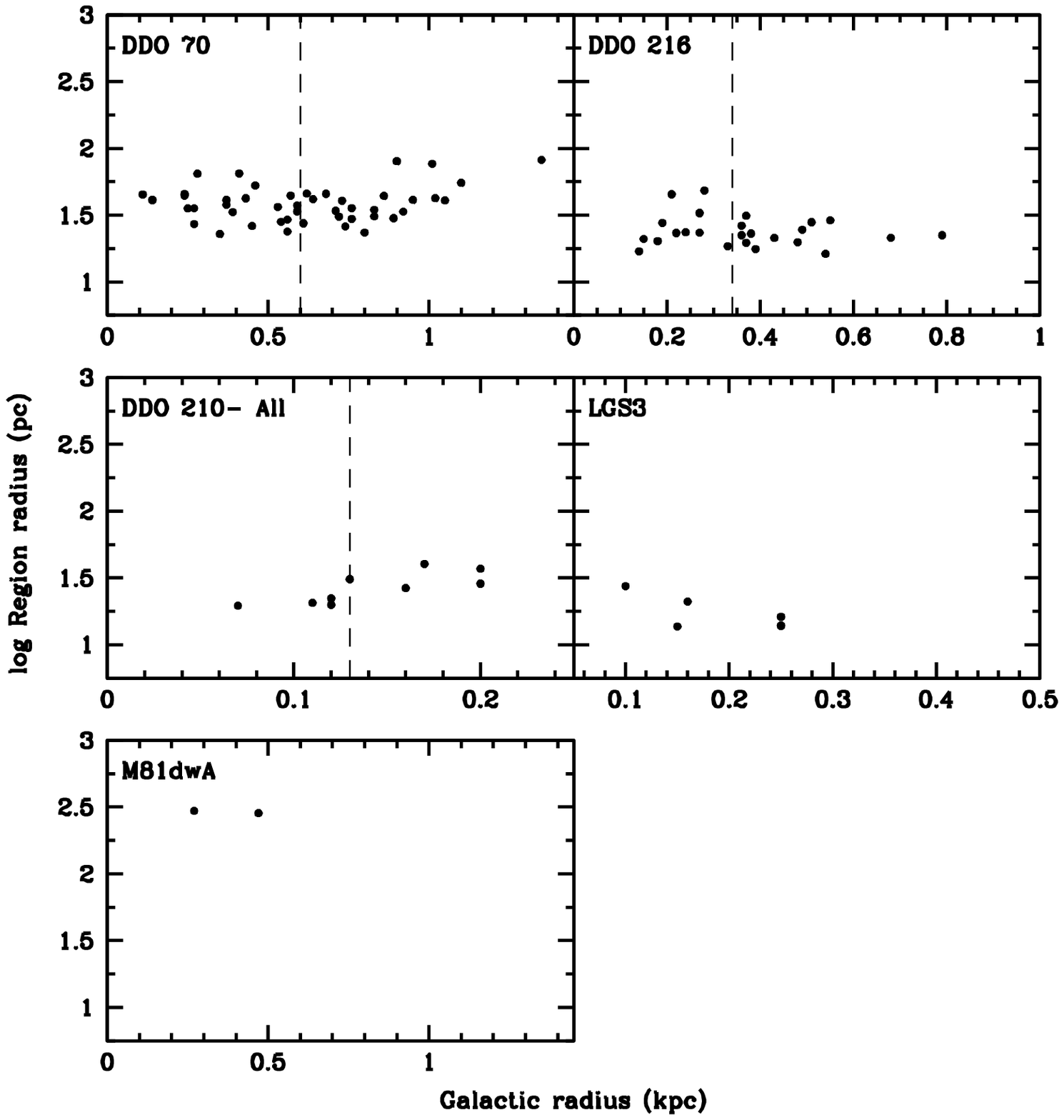} 
Figure \protect\ref{fig-radii} (continued)


\clearpage

\begin{figure}
\epsscale{1.0}
\plotone{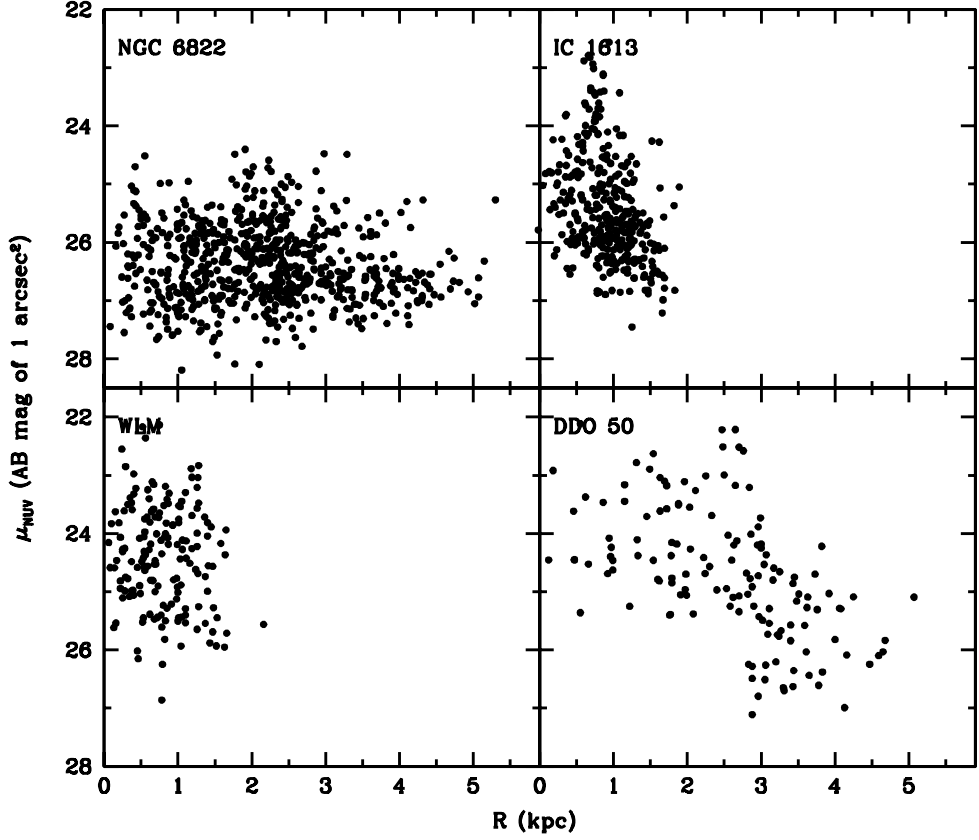} 
\caption{$NUV$ surface brightness
of each star-forming region in four galaxies, versus the distance from
the galactic center. The surface brightness decreases with distance for
DDO 50 partly because of a selection effect where regions have to be
brighter than the surrounding disk in order to be picked for the
survey.  This effect is also slightly present for IC 1613 and WLM, but
not obviously for NGC 6822. 
\label{fig-sbvsdist}}
\end{figure}

\clearpage

\begin{figure}
\epsscale{1.0} 
\plotone{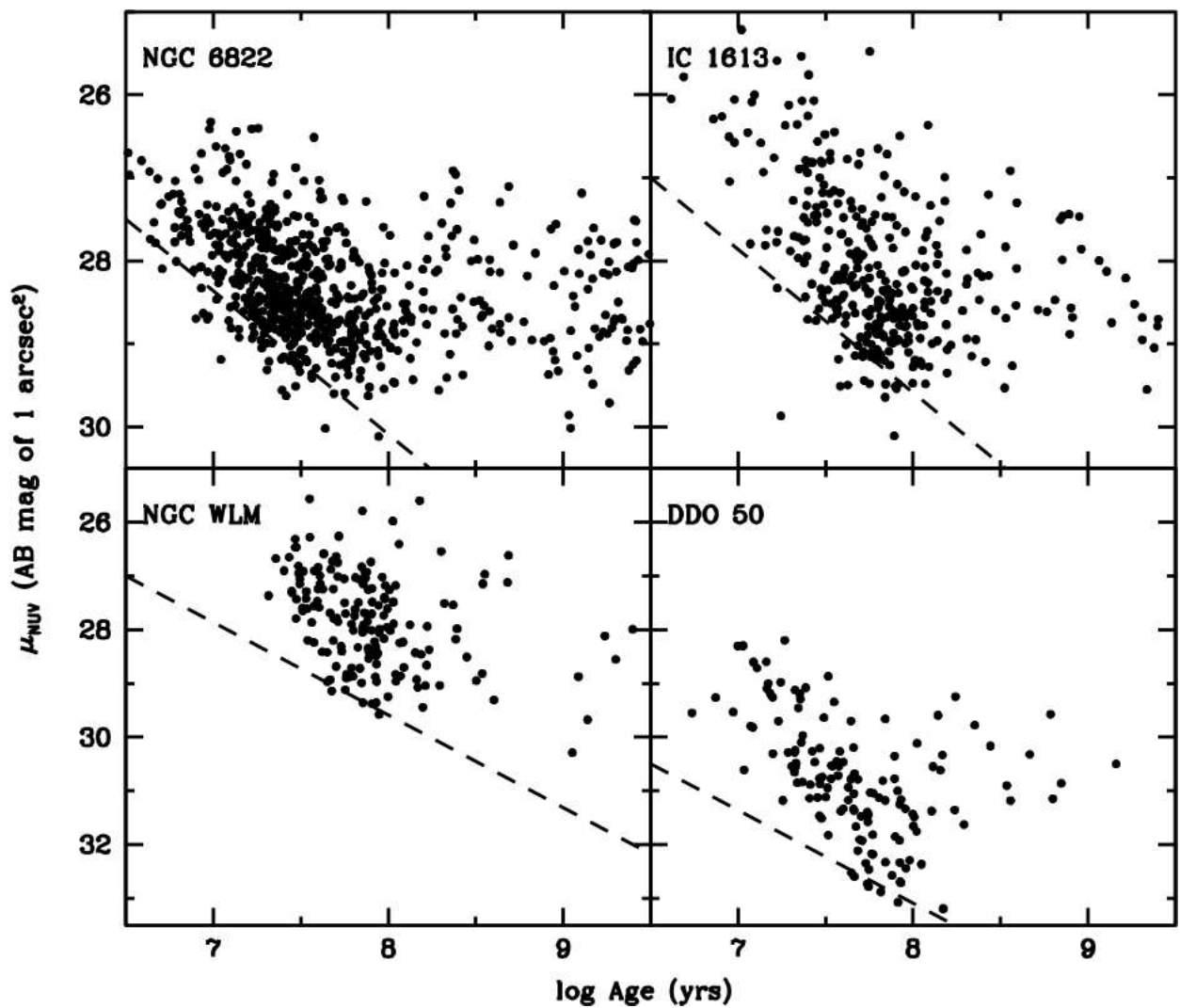} 
\caption{$NUV$ surface brightness of
each star-forming region in four galaxies, versus the log of the region
age. Dashed lines indicate the fading trend from the evolution of a
single stellar population model. Each galaxy has a lower limit to
surface brightness that depends on exposure time but not age. The top
left panel suggests a loss of surface brightness purely from stellar
evolution fading, while the other panels suggest some additional loss
because the distributions are steeper than the fading
trend.
\label{fig-regsb}}
\end{figure}

\clearpage

\begin{figure}
\epsscale{1.0} 
\plotone{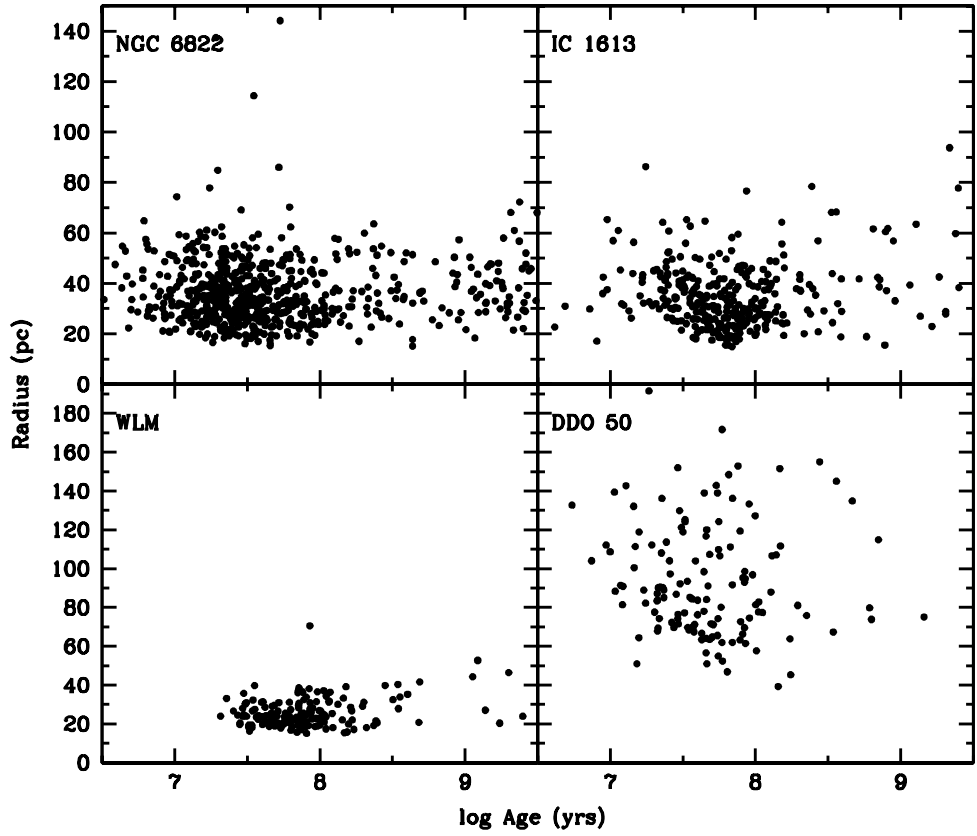} 
\caption{$NUV$ region radius
is shown versus $\log{\rm age}$. The regions have about the same size
for all ages, indicating no expansion of their visible outer
boundaries. 
\label{fig-radiusvsage}}
\end{figure}

\clearpage

\begin{figure}
\epsscale{1.0} 
\plotone{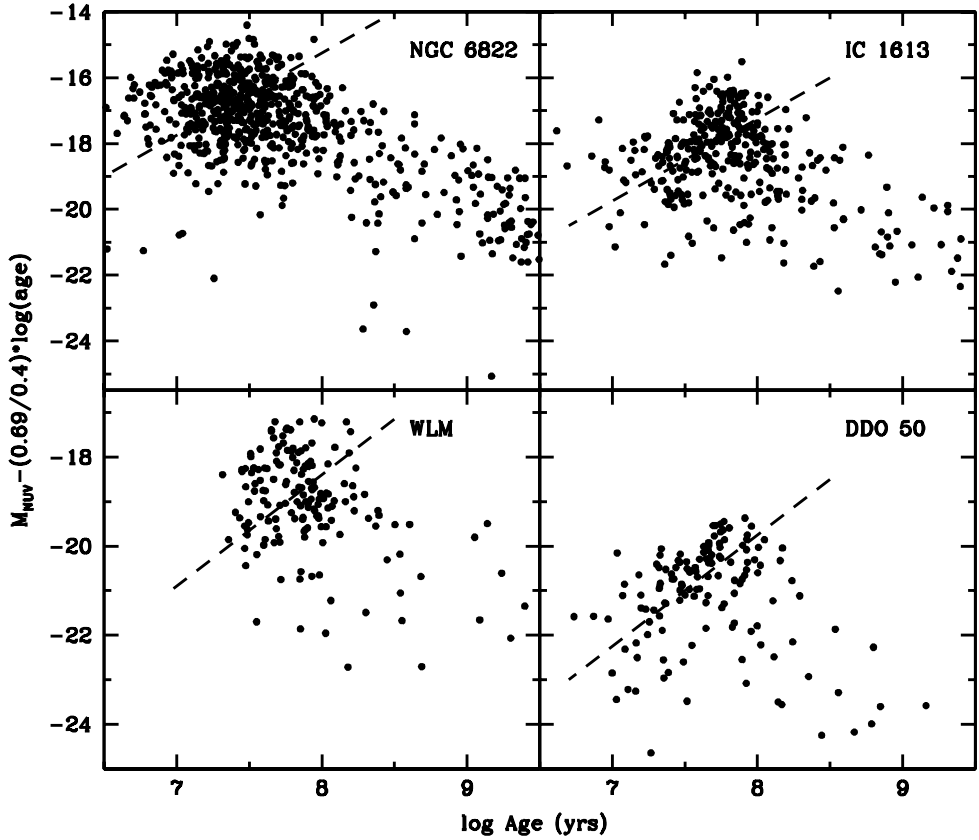} 
\caption{The fading-corrected
absolute magnitude is plotted versus the log of the region age. The
dashed line has a slope of $+2.5$ and traces the outline of dense
points, which show an upward trend. This trend suggests each region has
a mass that varies as $1/{\rm age}$.  
\label{fig-mnuvvsage}}
\end{figure}

\clearpage

\begin{figure}
\epsscale{1.0} 
\plotone{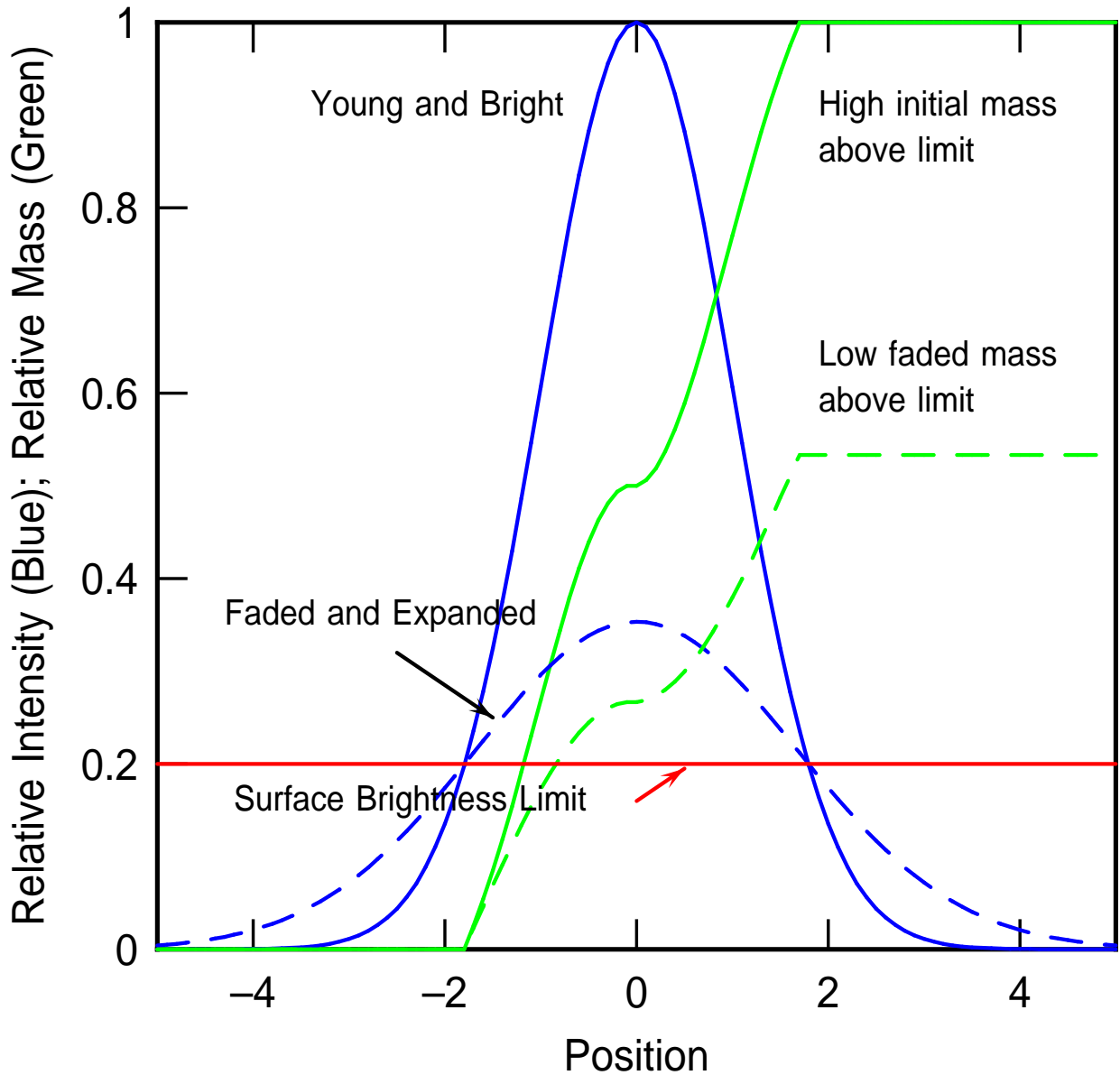} 
\caption{A schematic model
showing projected profiles (Gaussians) through two clusters with the
same total mass but different masses above the surface brightness
limit. The green rising curves are the cumulative masses above the
surface brightness limit, integrated from left to right over the
profiles. The mass above the surface brightness limit is smaller for
the fainter region, even though the total mass is the same. This model
illustrates how observable regions can appear to lose mass over and
above any effects from stellar evolution, evaporation, and disruption.
\label{fig-massloss_sblimit}}
\end{figure}


\clearpage

\begin{figure}
\epsscale{1.0}
\plotone{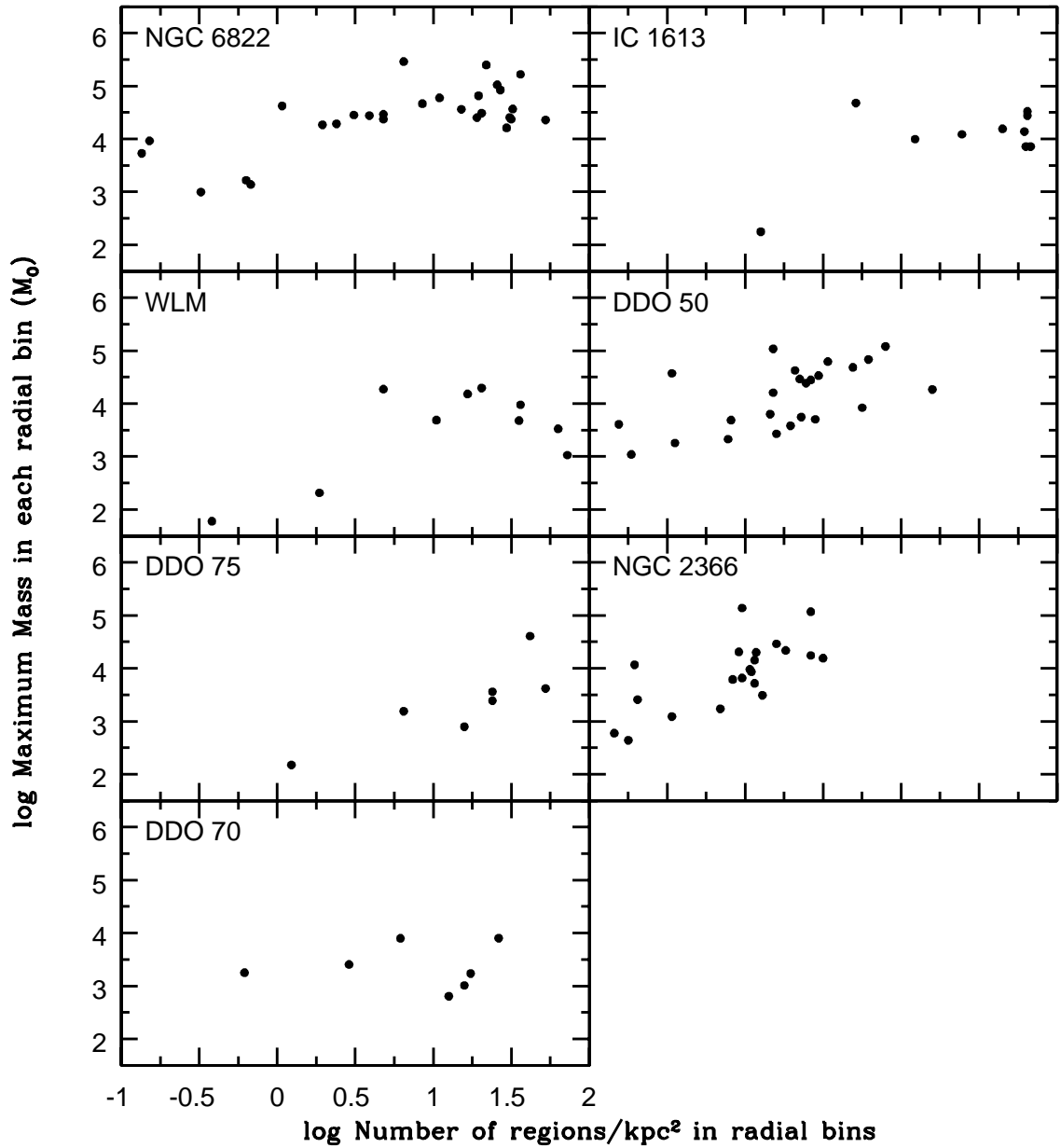}
\caption{Maximum cluster
mass in equal intervals of galactocentric radius plotted versus the
number of regions per unit area in those intervals. Bins are in 200 pc
increments of radius.
\label{fig-radnmaxmass}}
\end{figure}

\clearpage
\begin{figure}
\epsscale{1.0}
\plotone{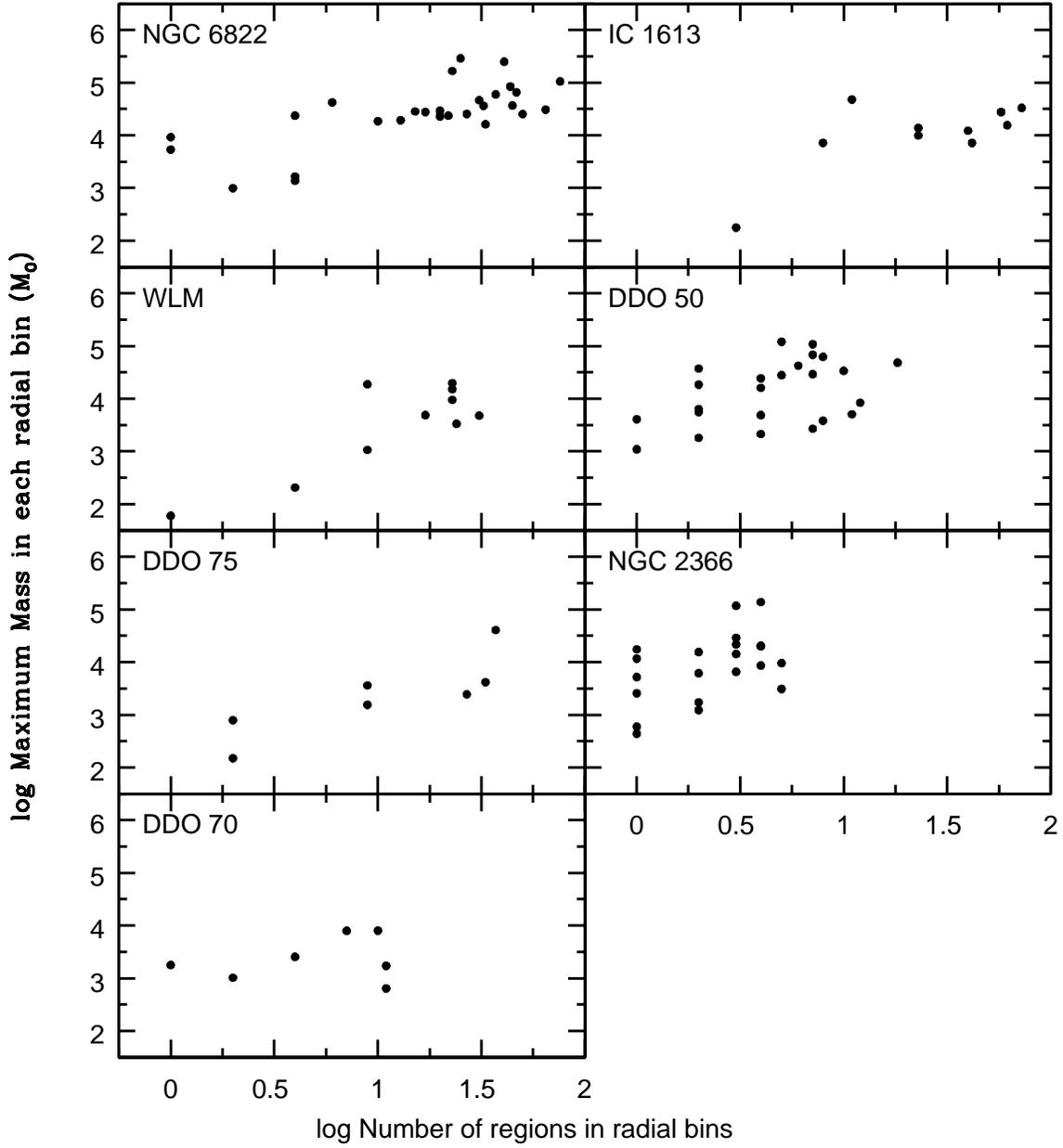}
\caption{Maximum cluster
mass in equal intervals of galactocentric radius plotted versus the
total number of regions in those intervals. Bins are in 200 pc
increments. The outer parts of most galaxies tend to have the smallest
numbers of regions. There is a trend from the size of sample effect
with a slope of unity on this diagram.
\label{fig-radnmaxmass2}}
\end{figure}

\clearpage

\begin{figure}
\epsscale{1.0}
\plotone{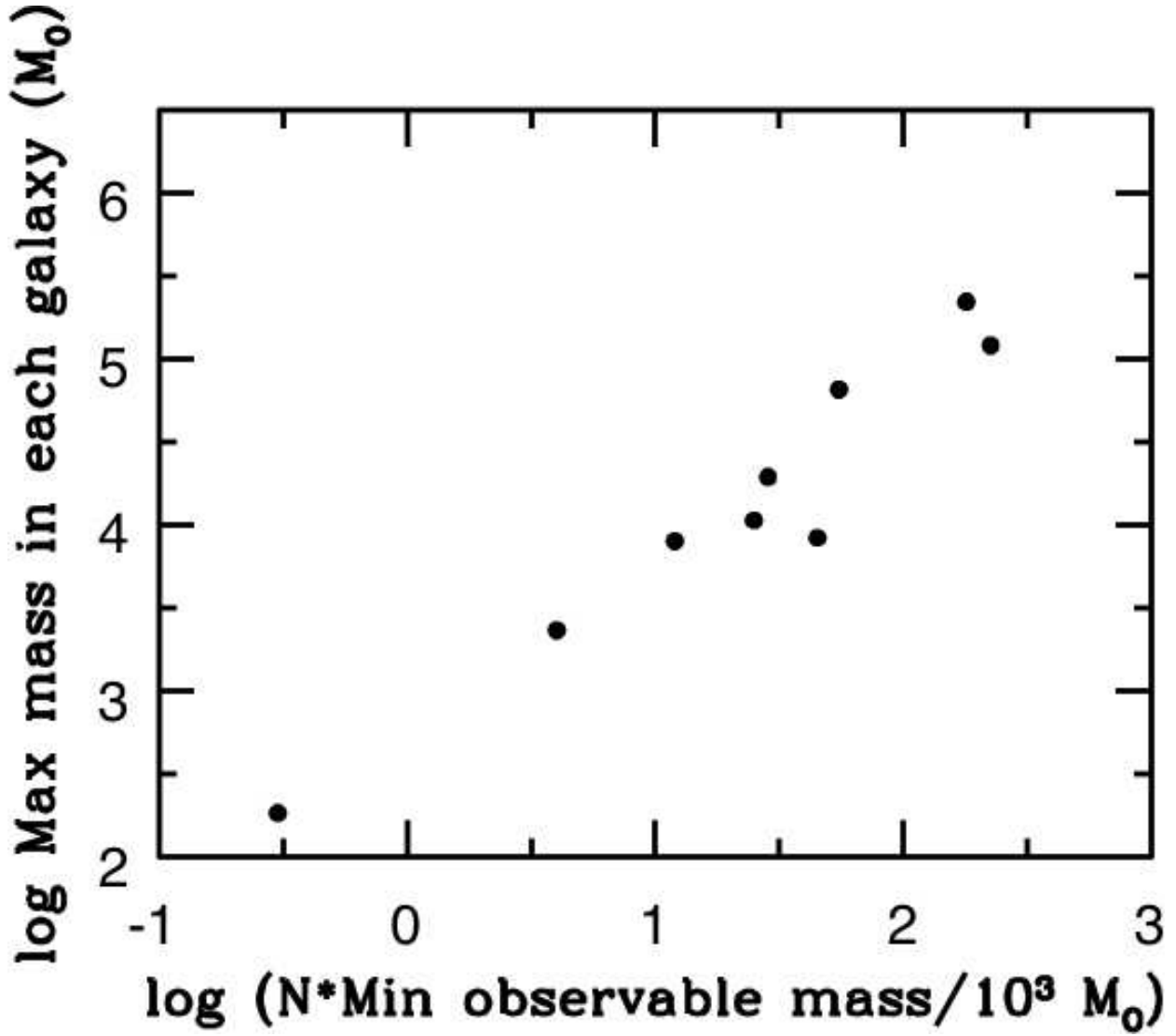}
\caption{Maximum region mass
in each galaxy, regardless of position or age, plotted versus the
total number of regions extrapolated to a mass of $10^3\;M_\odot$.  The
power law correlation with slope 1 follows from the size-of-sample
effect in a population of star-forming regions that have a mass
function $dN/dM\propto M^{-2}$. This correlation suggests that larger
and more active dwarf galaxies have more massive clusters entirely
because of sampling statistics, without any physical differences in the
detailed star formation mechanisms.
\label{fig-nmx}}
\end{figure}

\clearpage

\begin{figure}
\epsscale{1.0}
\plotone{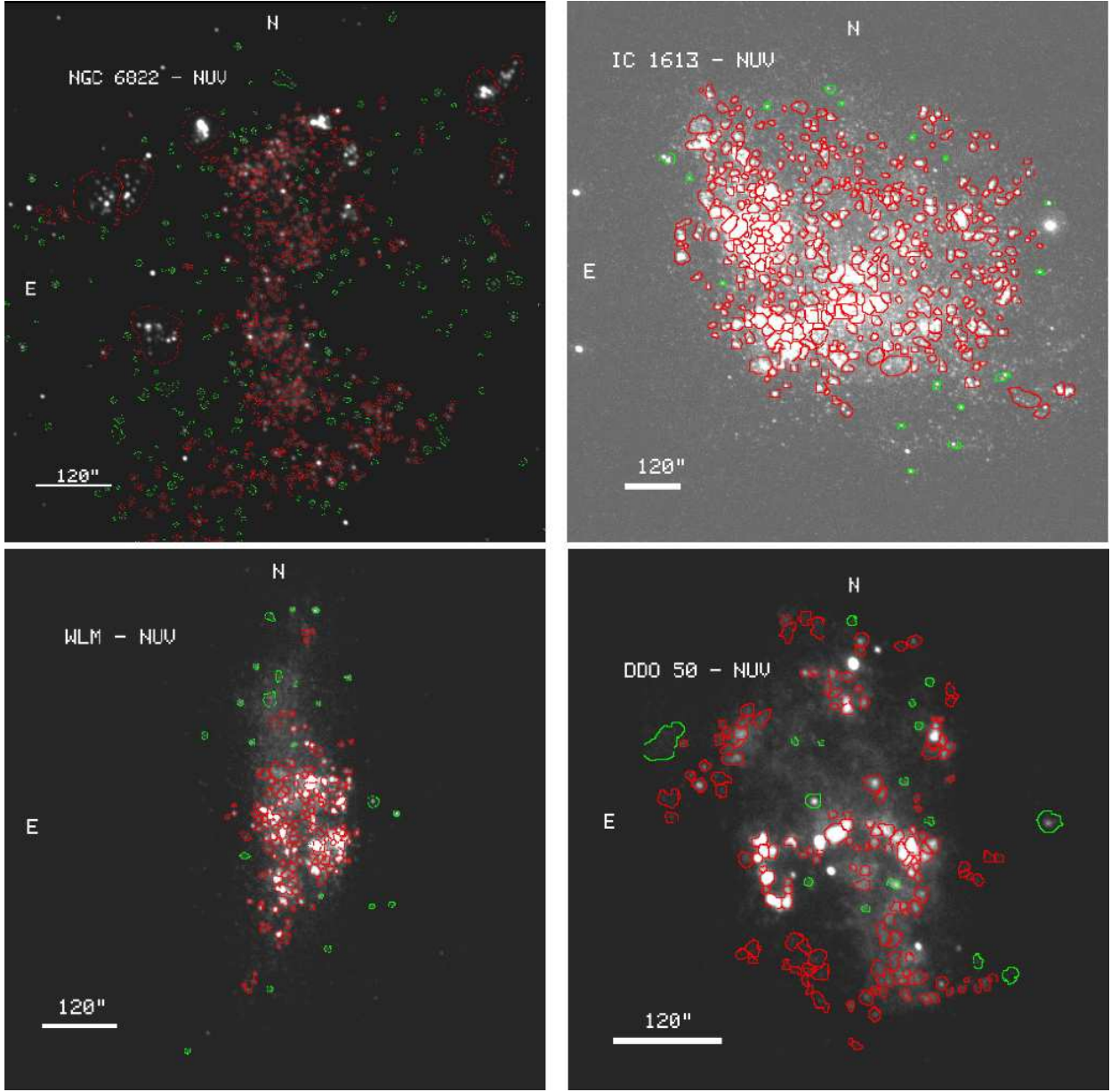}
\caption{$NUV$ images of each galaxy with individual star-forming regions outlined.
For regions outlined in green, the background was determined from a surrounding annulus.
For regions outlined in red, the background was determined by sampling areas around
the star-forming region that seemed to represent the underlying stellar disk.
\label{fig-images}}
\end{figure}

\clearpage

\plotone{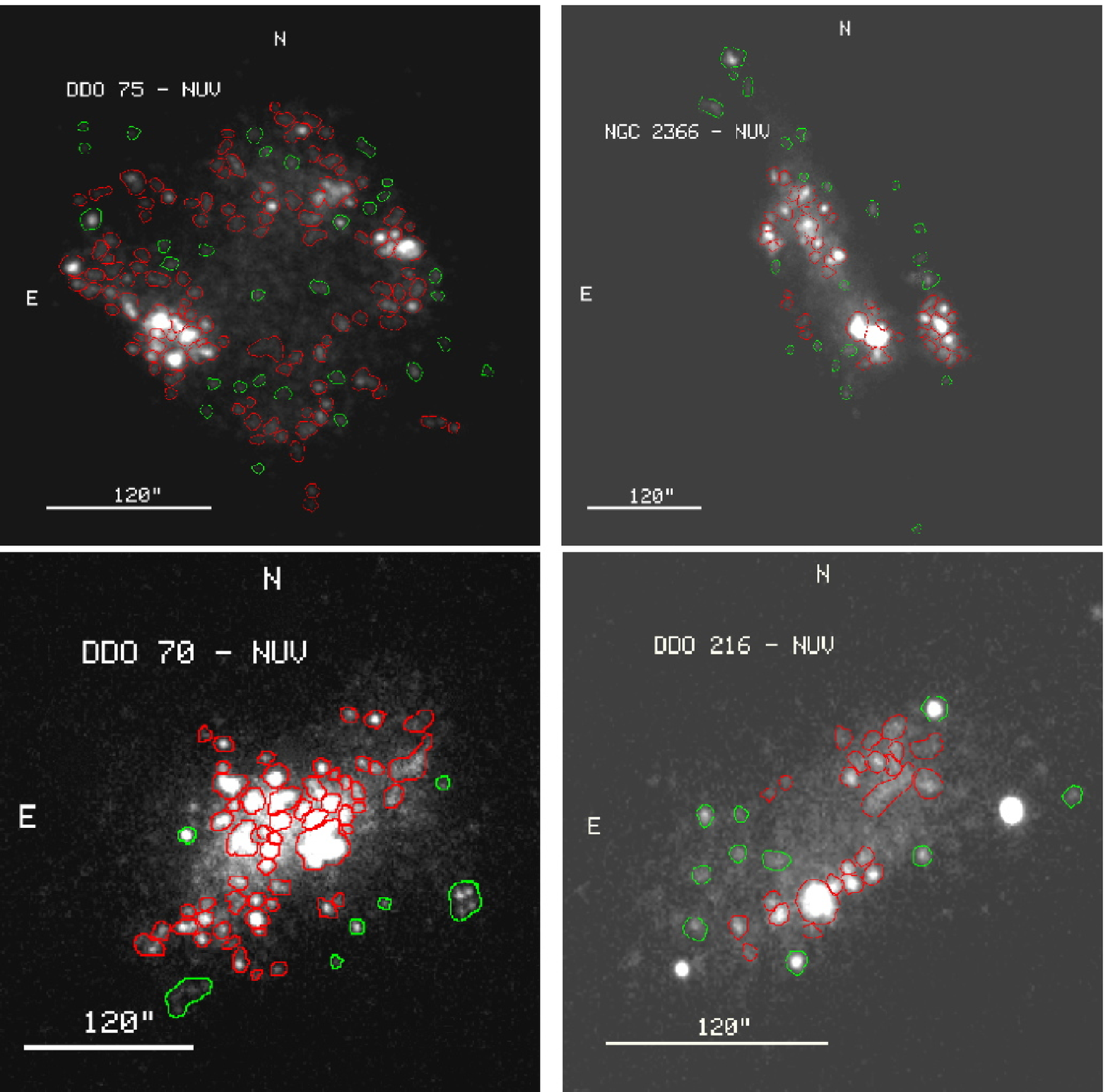}
Figure \ref{fig-images} (continued)

\plotone{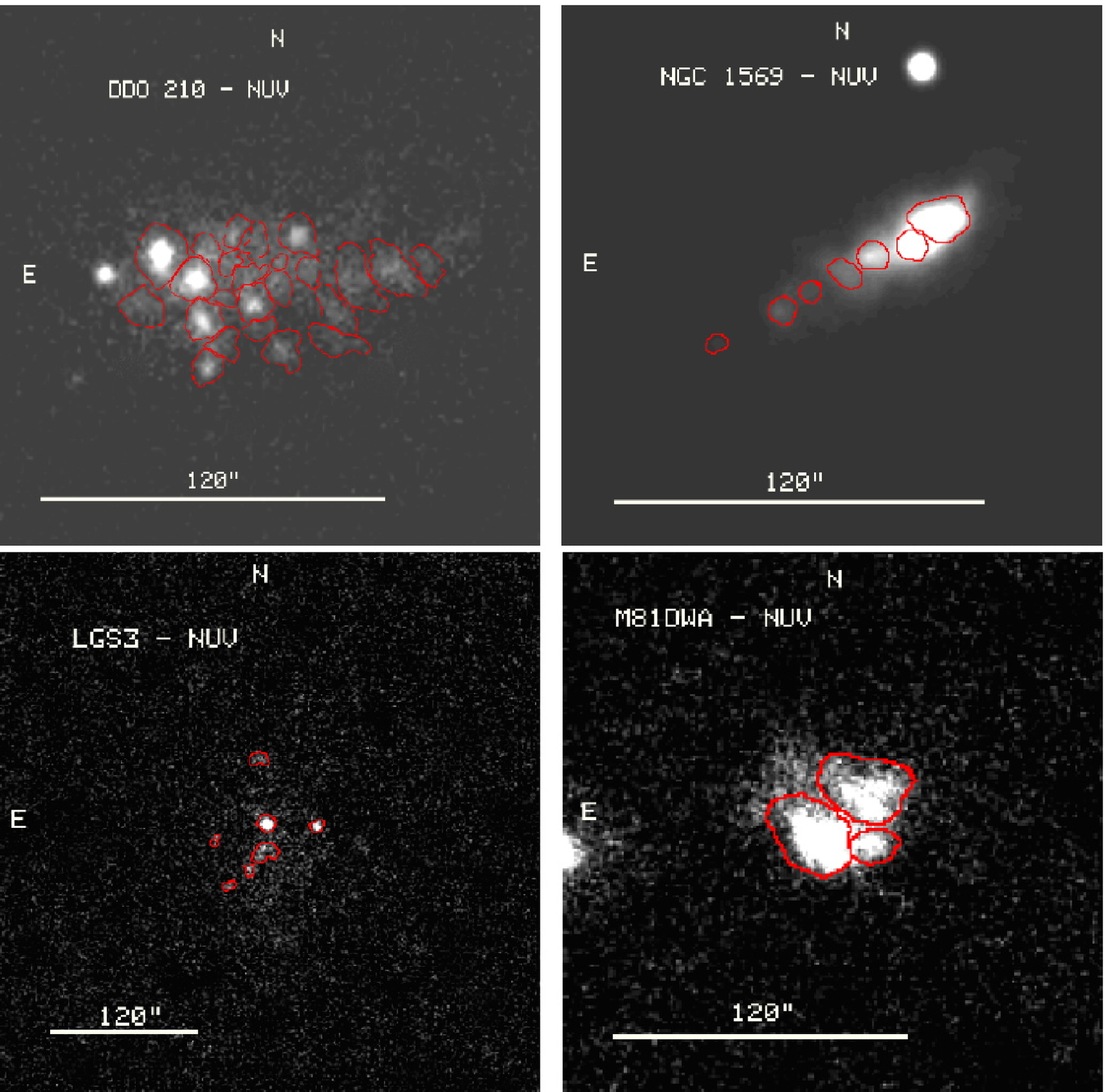}
Figure \ref{fig-images} (continued)

\clearpage

\begin{deluxetable}{llccrcrc}
\tabletypesize{\scriptsize}
\rotate
\tablenum{1}
\tablecolumns{8}
\tablewidth{0pt}
\tablecaption{Galaxy Sample\tablenotemark{a} \label{tab-sample}}
\tablehead{
\colhead{}
& \colhead{}
& \colhead{D}
& \colhead{}
& \colhead{}
& \colhead{$R_D$\tablenotemark{c}}
& \colhead{P.A.\tablenotemark{d}}
& \colhead{Incl.\tablenotemark{e}} \\
\colhead{Galaxy}
& \colhead{Other Names}
& \colhead{(Mpc)}
& \colhead{E($B-V$)$_f$\tablenotemark{b}}
& \colhead{M$_V$}
& \colhead{(arcmin)}
& \colhead{(deg)}
& \colhead{(deg)}
}
\startdata
DDO 50  \dotfill & PGC 23324,UGC 4305,Holmberg II,VIIZw223    & 3.4 & 0.02 & $-$16.6 &   1.11  &    18 & 46 \\
DDO 70  \dotfill & PGC 28913,UGC 5373,Sextans B               & 1.3 & 0.01 & $-$14.1 &   1.26  &    88 & 57 \\
DDO 75  \dotfill & PGC 29653,UGCA 205,Sextans A               & 1.3 & 0.02 & $-$13.9 & \nodata &    41 & 33 \\
DDO 210 \dotfill & PGC 065367,Aquarius dwarf                  & 0.9 & 0.04 & $-$10.9 &   0.63  & $-$85 & 66 \\
DDO 216 \dotfill & PGC 71538,UGC 12613,Peg DIG                & 0.9 & 0.02 & $-$13.3 &   1.68  & $-$58 & 69 \\
IC 1613 \dotfill & PGC 3844,UGC 668,DDO 8                     & 0.7 & 0.00 & $-$14.6 &   2.87  &    71 & 37 \\
LGS3    \dotfill & PGC 3792,Pisces dwarf                      & 0.6 & 0.04 &  $-$9.4 &   1.13  &  $-$3 & 64 \\
M81dwA  \dotfill & PGC 23521                                  & 3.6 & 0.02 & $-$11.7 & \nodata &    86 & 45 \\
NGC 2366\dotfill & PGC 21102,UGC 3851,DDO 42                  & 3.2 & 0.04 & $-$16.7 &   1.37  &    32 & 72 \\
NGC 6822\dotfill & PGC 63616,IC 4895,DDO 209,Barnard's Galaxy & 0.5 & 0.21 & $-$15.2 &   3.92  &    24 & 40 \\
WLM     \dotfill & PGC 143,UGCA 444,DDO 221                   & 1.0 & 0.02 & $-$14.4 &   1.97  &  $-$2 & 70 \\
\enddata
\tablenotetext{a}{Quantities taken from Hunter \& Elmegreen (2006).}
\tablenotetext{b}{Foreground reddening from Burstein \& Heiles (1984).}
\tablenotetext{c}{Disk scale length determined from $V$-band images.}
\tablenotetext{d}{Galactic position angle of the major axis determined from $V$-band images.}
\tablenotetext{e}{Inclination of galactic disk.}
\end{deluxetable}

\clearpage

\begin{deluxetable}{lccl}
\tablenum{2}
\tablecolumns{4}
\tablewidth{0pt}
\tablecaption{{\it GALEX} Observations \label{tab-obs}}
\tablehead{
\colhead{}
& \colhead{FUV Exp (s)}
& \colhead{NUV Exp (s)}
& \colhead{Tile Name}
}
\startdata
DDO 50  \dotfill & 1674 & 1521 & NGA\_HolmbergII  \\
DDO 70  \dotfill & 1076 &  966 & NGA\_SextansB \\
DDO 75  \dotfill & 1663 & 1512 & NGA\_SextansA \\
DDO 210 \dotfill & 1669 & 1472 & GI1\_047107\_DDO210 \\
DDO 216 \dotfill & 1557 & 2665 & MISDR2\_28664\_0746 \\
IC 1613 \dotfill & 1676 & 1494 & NGA\_IC16134 \\
LGS3    \dotfill & 1673 & 1487 & NGA\_LGS3 \\
M81dwA  \dotfill & 1674 & 1521 & NGA\_HolmbergII  \\
NGC 2366\dotfill & 2889 & 2617 & NGA\_NGC2366 \\
NGC 6822\dotfill & 4483 & 5132 & NGA\_NGC6822 \\
WLM     \dotfill & 1423 & 1294 & NGA\_WLM  \\
\enddata
\end{deluxetable}

\clearpage

\begin{deluxetable}{rrrrcrcrcrrrcrcrccr}
\rotate
\tabletypesize{\scriptsize}
\raggedright
\tablewidth{0pc}
\tablenum{3}
\tablecolumns{19}
\tablecaption{Region Photometry \label{tab-phot}}
\tablehead{
\colhead{}
&\colhead{\tiny{R.A.}}
&\colhead{\tiny{Decl.}}
&\colhead{\tiny{$R$\tablenotemark{a}}}
&\colhead{}
&\colhead{\tiny{$r_{reg}$\tablenotemark{b}}}
&\colhead{}
&\colhead{}
&\colhead{}
&\colhead{}
&\colhead{}
&\colhead{}
&\colhead{}
&\colhead{}
&\colhead{}
&\colhead{}
&\colhead{$\log M$\tablenotemark{d}}
&\colhead{$\log$ Age\tablenotemark{d}}
&\colhead{} \\
\colhead{\tiny{Region}}
&\colhead{\tiny{(J2000)}}
&\colhead{\tiny{(J2000)}}
&\colhead{\tiny{(kpc)}}
&\colhead{\tiny{$R/R_D$\tablenotemark{a}}}
&\colhead{\tiny{(pc)}}
&\colhead{\tiny{$M_{NUV}$\tablenotemark{c}}}
&\colhead{$\sigma$}
&\colhead{\tiny{$(FUV-NUV)_0$}}
&\colhead{$\sigma$}
&\colhead{\tiny{$M_V$}}
&\colhead{$\sigma$}
&\colhead{\tiny{$(B-V)_0$}}
&\colhead{$\sigma$}
&\colhead{\tiny{$(U-B)_0$}}
&\colhead{$\sigma$}
&\colhead{(M\protect\solar)}
&\colhead{(yrs)}
&\colhead{RMS\tablenotemark{d}}
}
\startdata
\cutinhead{DDO 50}
  1 &  8 19 09.6 &  70 45 19 &  1.96 &  1.78 &   108.1 &    -9.9 &  0.01 &  -0.08 &  0.03 &    -9.5 &   0.0 &  -0.01 &  0.02 &  -0.94 &  0.03 &  4.03 &  7.35 &  0.38 \\
  2 &  8 18 49.7 &  70 44 60 &  2.99 &  2.72 &    81.4 &    -8.6 &  0.02 &  -0.44 &  0.04 &    -8.5 &   0.0 &  -0.02 &  0.04 &  -1.00 &  0.05 &  3.55 &  7.08 &  0.38 \\
  3 &  8 19 12.1 &  70 43 07 &  0.55 &  0.50 &   191.6 &   -12.1 &  0.00 &  -0.17 &  0.01 &   -12.2 &   0.0 &   0.21 &  0.00 &  -1.14 &  0.00 &  5.08 &  7.27 &  0.55 \\
  4 &  8 19 27.8 &  70 42 21 &  2.65 &  2.41 &   139.4 &   -11.3 &  0.01 &  -0.22 &  0.01 &   -11.0 &   0.0 &   0.05 &  0.01 &  -1.20 &  0.01 &  4.53 &  7.03 &  0.55 \\
  5 &  8 19 28.6 &  70 42 30 &  2.65 &  2.40 &    64.4 &    -8.7 &  0.02 &  -0.15 &  0.04 &    -8.0 &   0.0 &  -0.05 &  0.04 &  -1.09 &  0.05 &  3.41 &  7.20 &  0.49 \\
\enddata
\tablecomments{Table \protect\ref{tab-phot} is published
in its entirety in the electronic edition
of the {\it Astronomical Journal}. A portion is shown here
for guidance regarding its form and content.}
\tablenotetext{a}{Distance of the star-forming region from the center of the
                  galaxy in the plane of the galaxy, using the galactic parameters given in
                  Table \protect\ref{tab-sample}.
                  $R_D$ is the disk scale length from Hunter \& Elmegreen (2006).}
\tablenotetext{b}{Radius of the star-forming region. The identifying regions are polygons, and the
                  radius is the square-root of the area of the region divided by $\pi$.}
\tablenotetext{c}{Absolute AB magnitude in the {\it NUV} filter.}
\tablenotetext{d}{Region stellar mass and age from the best fit model to the colors
                  and luminosity, and RMS.
                  Here RMS is the rms deviation between observed and modeled colors.}
\end{deluxetable}

\clearpage

\begin{deluxetable}{rrrrrrr}
\raggedright
\tablewidth{0pc}
\tablenum{4}
\tablecolumns{7}
\tablecaption{NGC 2366 Region JHK Photometry \label{tab-jhk}}
\tablehead{
\colhead{Region}
&\colhead{$M_{J}$}
&\colhead{$\sigma$}
&\colhead{$(J-H)_0$}
&\colhead{$\sigma$}
&\colhead{$H-K$}
&\colhead{$\sigma$}
}
\startdata
1 &   \nodata &  \nodata & \nodata &  \nodata & \nodata &  \nodata \\
2 &   \nodata &  \nodata & \nodata &  \nodata & \nodata &  \nodata \\
3 &   -9.597 &  1.187 &  1.159 &  2.222 &  0.156 &  4.851 \\
4 &   \nodata &  \nodata & \nodata &  \nodata & \nodata &  \nodata \\
5 &   -8.920 &  1.690 &  0.750 &  4.076 & \nodata &  \nodata \\
6 &   -8.980 &  1.558 &  0.548 &  4.309 & -0.382 & 18.097 \\
7 &  -10.637 &  0.465 &  0.906 &  1.014 &  0.217 &  2.236 \\
8 &   -9.613 &  1.093 &  0.803 &  2.546 &  0.028 &  6.839 \\
9 &   -8.834 & 13.849 & -3.127 & \nodata &  2.620 & 97.900 \\
10 & -10.708 &  0.580 &  0.617 &  1.531 &  0.523 &  2.727 \\
11 &  \nodata &  \nodata & \nodata &  \nodata & \nodata &  \nodata \\
12 &  -9.597 &  1.467 &  1.361 &  2.466 & -0.575 & 10.260 \\
13 &  -8.616 &  1.648 &  0.526 &  4.627 &  0.740 &  7.051 \\
14 &  -9.944 &  0.607 &  0.615 &  1.605 &  0.441 &  3.066 \\
15 &  -9.273 &  6.401 & -2.648 & 69.743 & \nodata &  \nodata \\
16 &  \nodata &  \nodata & \nodata &  \nodata & \nodata &  \nodata \\
17 &  \nodata &  \nodata & \nodata &  \nodata & \nodata &  \nodata \\
18 &  \nodata &  \nodata & \nodata &  \nodata & \nodata &  \nodata \\
19 &  \nodata &  \nodata & \nodata &  \nodata & \nodata &  \nodata \\
20 & -13.411 &  0.056 &  0.628 &  0.146 &  0.173 &  0.354 \\
21 & -11.129 &  0.258 &  0.443 &  0.768 &  0.287 &  1.723 \\
22 & -10.448 &  0.416 &  0.491 &  1.199 &  0.289 &  2.672 \\
23 &  -8.590 &  2.470 &  0.561 &  6.761 & \nodata &  \nodata \\
24 & -11.071 &  0.365 &  0.253 &  1.246 &  0.398 &  2.593 \\
25 &  -9.930 &  0.693 &  0.580 &  1.879 &  0.409 &  3.713 \\
26 &  -9.426 &  1.176 & -0.460 &  6.776 & \nodata &  \nodata \\
27 & -10.207 &  0.457 &  0.563 &  1.250 &  0.287 &  2.760 \\
28 &  -7.039 &  9.942 &  1.465 & 15.872 & \nodata &  \nodata \\
29 &  -6.648 & 10.252 &  0.668 & 26.066 & \nodata &  \nodata \\
30 &  -7.238 &  7.278 &  1.626 & 10.833 & -0.140 & 26.445 \\
31 &  \nodata &  \nodata & \nodata &  \nodata & \nodata &  \nodata \\
32 &  -9.510 &  1.560 & -0.031 &  6.561 & \nodata &  \nodata \\
33 & -10.935 &  0.407 &  0.534 &  1.137 &  0.151 &  2.854 \\
34 &  -9.537 &  0.989 &  0.485 &  2.862 &  0.077 &  7.761 \\
35 &  -9.494 &  1.458 & -0.258 &  7.256 & \nodata &  \nodata \\
36 &  -9.962 &  0.887 & -0.364 &  4.760 & -0.062 & 15.992 \\
37 & -11.469 &  0.300 &  0.298 &  0.992 &  0.119 &  2.648 \\
38 & -10.357 &  0.438 &  0.651 &  1.129 &  0.146 &  2.794 \\
39 &  -8.240 &  2.872 &  1.397 &  4.743 & -0.634 & 20.645 \\
40 &  \nodata &  \nodata & \nodata &  \nodata & \nodata &  \nodata \\
41 & -11.936 &  0.180 &  0.617 &  0.476 &  0.155 &  1.175 \\
42 &  \nodata &  \nodata & \nodata &  \nodata & \nodata &  \nodata \\
43 &  -8.874 &  2.827 & -1.226 & 28.369 & \nodata &  \nodata \\
44 &  -8.561 &  3.035 & -0.636 & 1\nodata &  1.616 & 20.540 \\
45 &  -7.926 &  4.552 & -1.098 & 41.742 &  2.620 & 36.823 \\
46 &  \nodata &  \nodata & \nodata &  \nodata & \nodata &  \nodata \\
47 &  \nodata &  \nodata & \nodata &  \nodata & \nodata &  \nodata \\
48 &  -7.859 &  3.774 &  0.508 & 10.698 &  0.058 & 29.730 \\
49 &  -9.369 &  1.381 &  1.259 &  2.449 &  0.571 &  3.609 \\
50 &  -7.974 &  4.131 &  0.808 &  9.603 & \nodata &  \nodata \\
51 &  -8.102 &  2.600 &  0.512 &  7.373 &  0.815 & 10.638 \\
52 &  \nodata &  \nodata & \nodata &  \nodata & \nodata &  \nodata \\
53 &  -9.785 &  1.041 &  0.100 &  3.976 & \nodata &  \nodata \\
54 & -11.279 &  0.329 &  0.652 &  0.847 &  0.185 &  2.020 \\
55 & -13.037 &  0.089 &  0.638 &  0.230 &  0.185 &  0.550 \\
56 & -13.763 &  0.046 &  0.282 &  0.156 &  0.446 &  0.312 \\
57 & -10.459 &  0.419 &  0.614 &  1.106 &  0.172 &  2.690 \\
58 &  \nodata &  \nodata & \nodata &  \nodata & \nodata &  \nodata \\
\enddata
\end{deluxetable}


\begin{deluxetable}{lrcrrcccc}
\tablenum{5}
\tablecolumns{9}
\tablewidth{0pt}
\tablecaption{Galactic Parameters \label{tab-param}}
\tablehead{
\colhead{}
& \colhead{$R_{H\alpha}$\tablenotemark{a}}
& \colhead{$R_{Br}$\tablenotemark{b}}
& \colhead{}
& \colhead{$R_{UV}$\tablenotemark{d}}
& \colhead{$R_{half}$\tablenotemark{e}}
& \colhead{}
& \colhead{}
& \colhead{$\Sigma_{HI} (R_{UV})$\tablenotemark{g}} \\
\colhead{Galaxy}
& \colhead{(kpc)}
& \colhead{(kpc)}
& \colhead{N$_{UV}$\tablenotemark{c}}
& \colhead{(kpc)}
& \colhead{(kpc)}
& \colhead{\% past $R_{H\alpha}$\tablenotemark{f}}
& \colhead{\% past $R_{Br}$\tablenotemark{f}}
& \colhead{(M\protect\solar\ pc$^{-2}$)}
}
\startdata
DDO 50  \dotfill &   4.02  & \nodata & 139 & 5.07 & 2.70 &    7    & \nodata & 3.6 \\
DDO 70  \dotfill &   1.23  & \nodata &  46 & 1.35 & 0.60 &    2    & \nodata & \nodata \\
DDO 75  \dotfill &   1.17  &  0.67   & 119 & 1.30 & 0.69 &      2   &    52  & 3.4 \\
DDO 210 \dotfill &   0.00  & \nodata &   9 & 0.20 & 0.13 & \nodata & \nodata & 4.4 \\
DDO 216 \dotfill &   0.34  &  1.42   &  25 & 0.79 & 0.36 &   56    &     0   & \nodata \\
IC 1613 \dotfill &   1.53  & \nodata & 342 & 1.89 & 0.91 &    5    & \nodata & 1.5 \\
LGS3    \dotfill &   0.00  & \nodata &   5 & 0.25 & 0.16 & \nodata & \nodata & \nodata \\
M81dwA  \dotfill &   0.00  &  0.42   &   2 & 0.47 & 0.37 & \nodata &    50   & \nodata \\
NGC 2366\dotfill &   5.25  & \nodata &  58 & 5.47 & 2.79 &    2    & \nodata & 2.2 \\
NGC 6822\dotfill &   1.65  & \nodata & 713 & 5.88 & 2.16 &   66    & \nodata & \nodata \\
WLM     \dotfill &   1.24  & \nodata & 165 & 2.16 & 0.78 &   19    & \nodata & 1.4 \\
\enddata
\tablenotetext{a}{Radius beyond which H$\alpha$ emission is no longer detected
(Hunter \& Elmegreen 2004). Galaxies with $R_{H\alpha}$ of zero have no detected
H$\alpha$ emission.}
\tablenotetext{b}{Radius where the $V$-band surface photometry changes slope (Hunter \& Elmegreen 2006).}
\tablenotetext{c}{Number of retained regions identified on the $NUV$ image.}
\tablenotetext{d}{Radius where last star-forming region is detected in the $NUV$ image.}
\tablenotetext{e}{Radius interior to which is found half of the star-forming regions identified on the $NUV$ image.}
\tablenotetext{f}{Percentage of identified UV regions found beyond the radius $R_{H\alpha}$  where
                 H$\alpha$ emission is no longer detected
                 and beyond the break radius $R_{Br}$ for those galaxies with breaks in their
                 broad-band surface brightness profiles.}
\tablenotetext{g}{HI surface density at $R_{UV}$.}
\end{deluxetable}

\end{document}